\documentclass[twocolumn]{emulateapj}
\usepackage{amstext}
\usepackage{apjfonts}
\usepackage{amsmath}
\usepackage{amssymb}
\usepackage{enumerate}
\usepackage{graphicx}
\usepackage{xspace}
\usepackage{aas_macros}
\usepackage[colorlinks=true,linkcolor=blue,citecolor=blue]{hyperref}

\newcommand{\dmde}{dM/dE}
\newcommand{\md}{\dot{M}\left(t\right)}
\newcommand{\mdpeak}{\dot{M}_{\rm peak}}
\newcommand{\tpeak}{t_{\rm peak}}
\newcommand{\ms}{M_{\ast}}
\newcommand{\rs}{R_{\ast}}
\newcommand{\rt}{r_{\rm t}}
\newcommand{\mh}{M_{\rm h}}
\newcommand{\msun}{M_{\odot}}
\newcommand{\ninf}{n_{\infty}}
\newcommand{\FLASH}{\texttt{FLASH}\xspace}
\newcommand{\lodato}{LKP\xspace}
\begin{document}

\shortauthors{Guillochon, Ramirez-Ruiz}

\title{Hydrodynamical Simulations to Determine the Feeding Rate of Black Holes by the Tidal Disruption of Stars: The Importance of the Impact Parameter and Stellar Structure}

\author{James Guillochon and Enrico Ramirez-Ruiz \altaffilmark{1}}
\altaffiltext{1}{Department of Astronomy and
  Astrophysics, University of California, Santa Cruz, CA
  95064}
   
\begin{abstract} 
The disruption of stars by supermassive black holes has been linked to more than a dozen flares in the cores of galaxies out to redshift $z \sim 0.4$. Modeling these flares properly requires a prediction of the rate of mass return to the black hole after a disruption. Through hydrodynamical simulation, we show that aside from the full disruption of a solar mass star at the exact limit where the star is destroyed, the common assumptions used to estimate $\md$, the rate of mass return to the black hole, are largely invalid. While the analytical approximation to tidal disruption predicts that the least-centrally concentrated stars and the deepest encounters should have more quickly-peaked flares, we find that the most-centrally concentrated stars have the quickest-peaking flares, and the trend between the time of peak and the impact parameter for deeply-penetrating encounters reverses beyond the critical distance at which the star is completely destroyed. We also show that the most-centrally concentrated stars produced a characteristic drop in $\md$ shortly after peak when a star is only partially disrupted, with the power law index $n$ being as extreme as -4 in the months immediately following the peak of a flare. Additionally, we find that $n$ asymptotes to $\simeq -2.2$ for both low- and high-mass stars for approximately half of all stellar disruptions. Both of these results are significantly steeper than the typically assumed $n = -5/3$. As these precipitous decay rates are only seen for events in which a stellar core survives the disruption, they can be used to determine if an observed tidal disruption flare produced a surviving remnant. We provide fitting formulae for four fundamental quantities of tidal disruption as functions of the star's distance to the black hole at pericenter and its stellar structure: The total mass lost, the time of peak, the accretion rate at peak, and the power-law index shortly after peak. These results should be taken into consideration when flares arising from tidal disruptions are modeled.
\end{abstract}


\section{Introduction}

Supermassive black holes (SMBHs) have been found to reside at the centers of most galaxies. These black holes are orbited by a cluster of stars that interact with one another gravitationally through stochastic encounters. Occasionally, an encounter will shift a star onto an orbit that takes it within its tidal radius, defined as the distance at which the black hole's tidal forces would overcome the star's self-gravity at its surface \citep{Frank:1976tg}. A fraction of the star's mass then becomes bound to the black hole, and proceeds to fall back towards the star's original pericenter, eventually forming an accretion disk that results in a luminous flare with a luminosity comparable to the Eddington luminosity.

The standard model of tidal disruption presumes that the star is completely destroyed, resulting in approximately half of the star's original mass falling back onto the black hole, with the debris possessing a variety of orbital periods resulting from a spread of orbital energy that is ``frozen in'' at pericenter. First described in \cite{Rees:1988ei}, the rate of fallback has been estimated both through increasingly sophisticated numerical simulations and analytical models. While previous results have provided reasonable models for the fallback resulting from the complete disruptions of stars at the tidal radius $\rt = \rs (\mh/\ms)^{1/3}$, where $\ms$ and $\rs$ are the mass and radius of the star and $\mh$ is the mass of the black hole, they completely neglect partial stellar disruptions, in which a stellar core survives the encounter and only a fraction of the star's mass becomes immediately bound to the black hole. These events are likely to be much more common than their complete disruption counterparts, both for the reason that the rate of encounters interior to the pericenter distance $r_{\rm p}$ scales as $r_{\rm p}$ \citep{Hills:1988br}, and also that the disrupted star may return on subsequent orbits and be subject to disruption and/or further tidal dissipation. Additionally, many previous studies have focused on stars of a single structural profile, usually selected to match the familiar profile of our own Sun. However, standard stellar mass functions predict that low-mass main sequence (MS) stars are more common \citep[e.g.][]{Kroupa:1993tm}, and thus may contribute significantly to the overall disruption rate. These stars are significantly less centrally concentrated than their solar mass brethren.

The dynamics of stellar tidal disruption have been modeled by many authors using simple analytical arguments \citep{Rees:1988ei,Phinney:1989uw,Lodato:2009iba,Kasen:2010ci}, increasingly complex dynamical models \citep{Luminet:1985wz,Carter:1985ti,Luminet:1986ch,Diener:1995ui,Ivanov:2001fva}, and hydrodymical simulations utilizing either an Eulerian \citep{Evans:1989jua,Khokhlov:1993cu,Khokhlov:1993bj,Diener:1997kw,Guillochon:2009di} or Lagrangian \citep{Nolthenius:1982dn,Bicknell:1983dn,Laguna:1993cf,Kobayashi:2004kq,Rosswog:2008gc,Lodato:2009iba,RamirezRuiz:2009gw,Rosswog:2009gg,Antonini:2011ia} approaches. Very few of these studies have presented the effect varying $r_{\rm p}$ on the amount of mass lost by the star, $\Delta M$, or the effect on $\md$, the rate at which the mass liberated from the star returns to pericenter. Given that the viscous time is expected to be significantly shorter than the period of the returning debris, this $\md$ is expected to track the luminosity $L(t)$ closely. As the number of observed disruptions increases, and as the cadence and quality of data improves, it becomes increasingly more important to improve models of $\md$ for disruptions of all kinds.

In this paper, we present the results of 43 hydrodynamical simulations at high-resolution representing the disruption of both low-mass and high-mass MS stars. This provides, for the first time, a complete picture of the feeding of SMBHs by the disruption of MS stars. While the expected trend of smaller mass accretion rates for progressively more grazing encounters is reproduced, our study reveals several surprises on how disruptions work, particularly on the effect of stellar structure and how the fallback rate scales for both grazing and deep encounters. Contrary to what is expected from the freezing model, in which only the distribution of mass at pericenter is considered, the non-linear response of the star to the tidal field is found to play a crucial role in determining $\md$. Our simulations show that the simple models previously employed to predict the rate of fallback do not capture the full dynamics of the problem, and are only appropriate for anything other than the full disruption at exactly the tidal radius.

We find that the decay rate of $\md$ does not settle to a constant  value until a few months after the disruption for all disruptions by black holes with $\mh > 10^{6} \msun$, implying that the range of characteristic decay rates used to identify tidal disruption flares should be widened to include events that may not follow the fiducial $t^{-5/3}$ decay rate. For partial disruptions, the decay rate at a few years after the disruption depends crucially on the hydrodynamical evolution of the debris stream. This means that simulations must cover more than a few stellar dynamical timescales after the disruption, with the final functional form of $\md$ not being established until the star is many hundreds of tidal radii away from the black hole. And while we do find that there are differences between the fallback functions calculated for the disruptions of profiles characteristic of low- and high-mass stars, the mass-radius relationship of MS stars results in a family of fallback curves that are difficult to distinguish from one another for stars of  $0.3 \msun \gtrsim \ms \gtrsim 1.0 \msun$ without considering secondary features related to the shape of the fallback curves themselves, such as the decay rate of $\md$, characterized by a time-dependent power-law index $n(t)$.

This paper is organized as follows. We describe our numerical method, initial models, and method for calculating $\Delta M$ and $\md$ in Section \ref{sec:method}. The results of these simulations and how they improve our understanding of stellar tidal disruptions is described in Section \ref{sec:results}. A discussion of the general trends and their effect on the observable features of tidal disruptions is presented in Section \ref{sec:discussion}. Finally, we provide fitting formula to four characteristic variables describing disruptions of stellar profiles characteristic of low- and high-mass stars in Appendix \ref{sec:appendix}.

\section{Method}\label{sec:method}

Our simulations of tidal disruption are performed in \FLASH \citep{Fryxell:2000em}, an adaptive-mesh grid-based hydrodynamics code which includes self-gravity. The standard hydrodynamical equations are solved using the directionally split piecewise-parabolic approach \citep{Colella:1984wg} as provided by the \FLASH software, which has a small numerical viscosity and diffusivity as compared to the available unsplit solvers Therefore, the principle source of entropy generation is via shocks, if they are present. The solution to the Riemann problem is sensitive to the velocity of the frame in which the problem is solved, and poor solutions can be returned in regions where the velocity of the frame is many times larger than the sound speed \citep{Tasker:2008cu,Robertson:2010bk,Springel:2010hx}. As stars that are disrupted by SMBHs are traveling at a fraction of the speed of light $c$, we perform our simulations in the rest-frame of the star where the velocities are $\sim \sqrt{c_{\rm s} v_{\rm p}}$, where $c_{\rm s}$ is the sound speed within the star and $v_{\rm p}$ is the velocity at pericenter.

Our method is very similar to what is presented in \cite{Guillochon:2011be}, except that we now utilize version 4.0 of the \FLASH software, which has a greatly improved multipole gravity solver\footnote{See http://flash.uchicago.edu for details}. A key parameter of the multipole gravity solver is the maximum angular number of the multipole expansion $l_{\rm m}$. A test simulation setting $l_{\rm m} = 10$ showed multiple recollapse points for a nearly-complete disruption, which is not expected to occur in disrupted stars (as described in Section \ref{subsec:survival}). We suspected this behavior was a consequence of the large aspect ratio of the debris stream, which results in gravity being under-resolved if the multipole expansion is truncated at small $l_{\rm m}$. With $l_{\rm m}$ set to 20, only a single recollapse occurs, as is expected. For $l_{\rm m} = 40$, we found no significantly difference in any quantities of interest as compared to $l_{\rm m} = 20$, except for cases in which a very small remnant survives the disruption. In these cases, the error is in the mass of the surviving object, which is difficult to resolve for marginally surviving stars (see Section \ref{subsec:survival}). The results presented in this work all use $l_{\rm m} = 20$ for optimal runtime efficiency.

Additionally, we add a truncation density parameter $\rho_{\rm damp}$ that is set to $10^{-18}$ g cm$^{-1}$, a factor of 10 larger than the fluff density. Material with density less than this value is not included in the multipole expansion, nor are any gravitational forces applied to this material. This is necessary as the domain is extremely large as compared to the initial star, and so even $10^{-11} \msun$ of material can introduce a significant error in the calculated position of the center of mass. We also only include material with a density greater than 10\% of the star's original peak density when calculating the location to use as the multipole expansion point; however all material with density greater than $\rho_{\rm damp}$ is included when calculating the magnitude of each of the multipole terms. This is done because the total center of mass does not always spatially coincide with the peak density, which can result in a multipole expansion that is a poor representation of the underlying density distribution.

\begin{figure*}[t]
\centering\includegraphics[width=0.9\linewidth,clip=true]{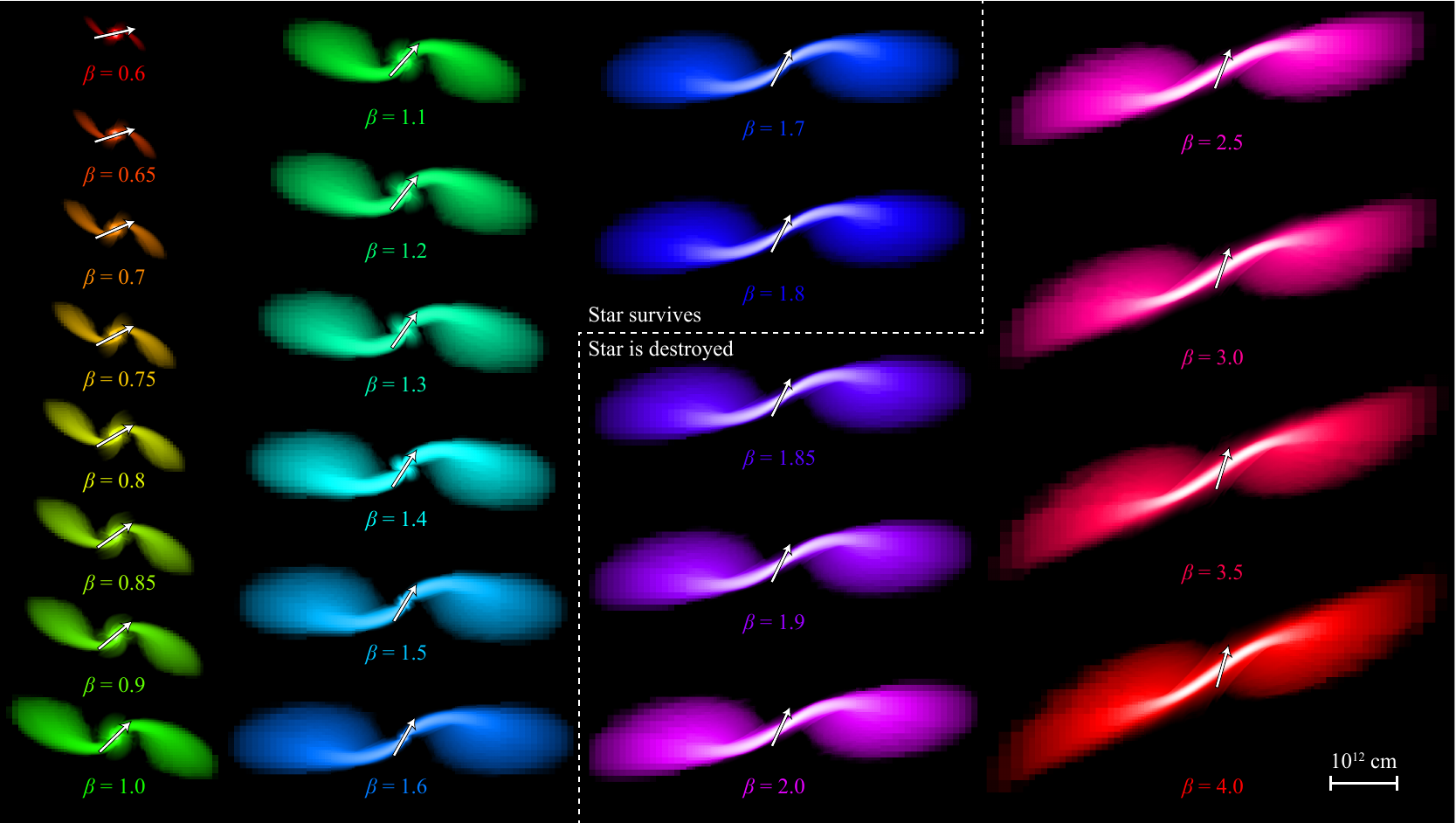}
\caption{Snapshots of the density $\log \rho$ for all $\gamma = 4/3$ simulations at $t = 4 \times 10^{4}$ s after the start of each simulation, with white corresponding to the maximum density and black corresponding to $10^{-6}$ g cm$^{-1}$. Each snapshot is labeled with the ratio of the tidal radius to the pericenter distance $\beta$. The white arrows indicate the angle of the velocity vector at the time of each snapshot. The white dashed line separates simulations in which a core survives the encounter; although not visible here, recollapse does occur for the $\beta =$ 1.7 and 1.8 simulations, but for $t > 4 \times 10^{4}$ s.}
\label{fig:disrupt-beta-4-3}
\end{figure*}

\subsection{Parameter Study}

Ignoring general relativistic effects and stellar rotation, it may seem that a complete study of tidal disruptions would require an exhaustive study of the various combinations of six parameters: $\ms$, $\rs$, $\mh$, the orbital eccentricity $e$, the polytropic index $\gamma$, and $\beta \equiv \rt/r_{\rm p}$. As an exhaustive search of a six-dimensional parameter space is prohibitive, we wish to reduce the number of free parameters to a more manageable number. For fixed $\beta$, both $r_{\rm p}$ and $v_{\rm p}$ scale as $\mh^{1/2}$, and thus the pericenter crossing time $t_{\rm p}$ is independent of $\mh$. Additionally, as the mass ratio approaches infinity, the asymmetry of the tidal field becomes progressively less important as $\rs \ll \rt$, with the difference in the strength of the tidal field at pericenter between the near-side and the far-side for a $10^{6}:1$ encounter being \(\simeq 3\%\) \citep{Guillochon:2011be}. And as most of the stars that are scattered into disruptive orbits originate from the sphere of influence or beyond \citep{Magorrian:1999vz,Wang:2004jy}, the orbital eccentricity of almost all disrupted stars is approximately unity\footnote{See \citep{Madigan:2011ej} for a discussion of resonant relation processes that may produce a different distribution of eccentricities for stellar disruptions.}.

\begin{figure*}[t]
\centering\includegraphics[width=0.9\linewidth,clip=true]{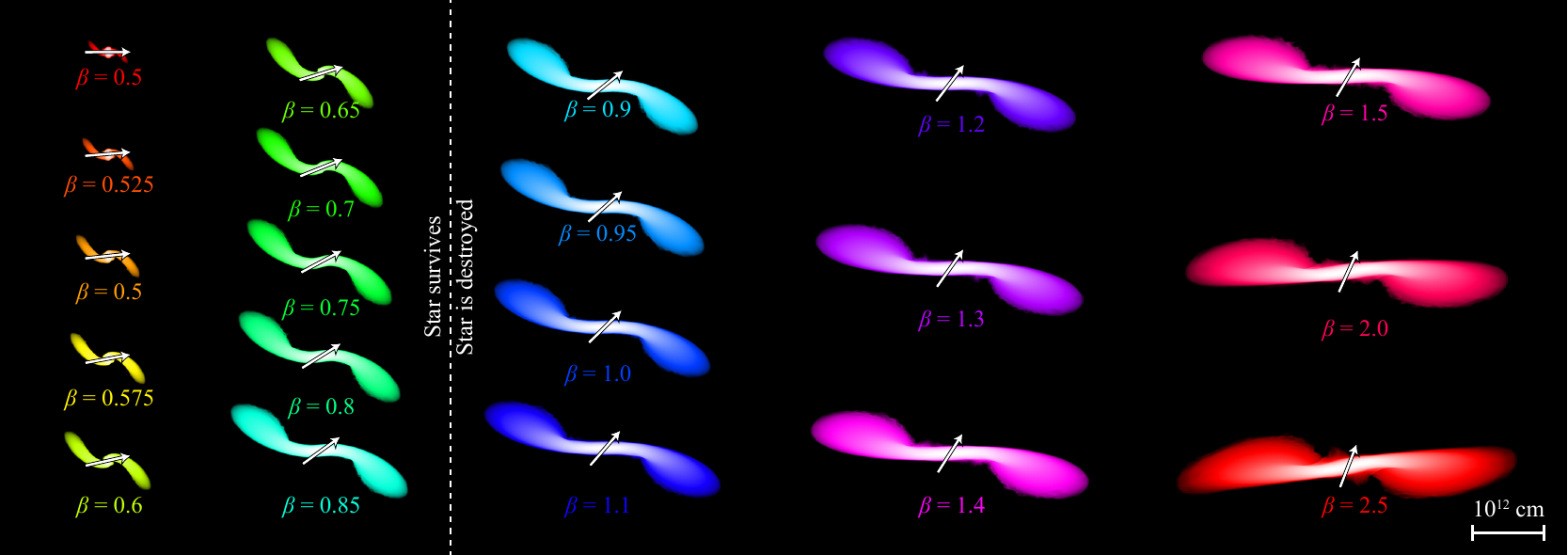}
\caption{Snapshots of $\log \rho$ for all $\gamma = 5/3$ simulations at $t = 4 \times 10^{4}$ s, colors and annotations are the same as in Figure \ref{fig:disrupt-beta-4-3}. Although not visible here, recollapse does occur for the $\beta =$ 0.75 -- 0.85 simulations, but for $t > 4 \times 10^{4}$ s.}
\label{fig:disrupt-beta-5-3}
\end{figure*}

It then follows that as the ratio of the time of the encounter $t_{\rm p}$ to the star's dynamical time $t_{\rm dyn} = \sqrt{R_{\ast}^{3}/GM_{\ast}}$, shape of the orbit (which depends on $e$), and asymmetry of the tides are nearly identical for all encounters of interest for a fixed $\beta$, the tidal force applied to the star as a function of time is independent of $\mh$, $e$, $\ms$, and $\rs$. Thus, we find that the vast majority of stellar disruptions by SMBHs can be described by just two parameters: $\beta$ and $\gamma$, with all other parameters obeying simple scaling relations. While previous numerical studies have considered the effect of varying $\gamma$ on $\md$\citep{Rosswog:2009gg,Lodato:2009iba,RamirezRuiz:2009gw}, the present work is the first to explore the effect of varying $\beta$ on $\md$ in cases ranging from no mass loss to deeply penetrating encounters\footnote{Note that \cite{Laguna:1993cf} do present $\md$ from low-resolution simulations for three different $\beta$ values, two of which are very deeply penetrating ($\beta \geq 5$).}.

To explore this reduced but physically motivated parameter space, we run a series of simulations assuming $\ms = \msun$ and $\mh = 10^{6} \msun$. Our stars are constructed as polytropes, with the polytropic $\gamma$ being set to either 5/3 or 4/3, representative of both low- and high-mass stars, respectively. During the simulation, the stars are evolved hydrodynamically according to a $\Gamma = 5/3$ equation of state, with the difference between $\gamma$ and $\Gamma$ for high-mass stars being a consequence of the radiation transfer within the star \citep{Chandrasekhar:1939vw}. These one-dimensional profiles are then mapped to the three-dimensional grid, with initially uniform refinement across the star. The star is then relaxed for $10^{4}$ s at the center of a cubical domain, which is $4 \times 10^{14}$ cm on a side. The domain is initially composed of a single $8^{3}$ block, which is then bisected into smaller $8^{3}$ blocks as many as 15 times, resulting in a minimum cell size of $3 \times 10^{9}$ cm, or approximately 2\% of the star's original diameter. Our refinement criteria is solely dependent on the density relative to the initial central density, with a factor of two reduction in resolution for each factor of a hundred in density. Regions within the simulation that are within 1\% of the peak density are always maximally refined.

We ran simulations for 23 different impact parameters $\beta \equiv \rt/r_{\rm p}$ ranging from 0.6 to 4.0 for $\gamma = 4/3$, and 20 different $\beta$ ranging from 0.5 to 2.5 for $\gamma = 5/3$. Two additional simulations were run at $\beta = 0.5$ for $\gamma = 4/3$ and $\beta = 0.45$ for $\gamma = 5/3$; as less than $10^{-6} \msun$ is observed to be removed from the stars in these two borderline cases, we conlude that no mass is lost for values of $\beta$ less than the above quoted ranges in $\beta$. Snapshots from each simulation recorded shortly after pericenter are shown in Figures \ref{fig:disrupt-beta-4-3} and \ref{fig:disrupt-beta-5-3}.

\subsection{Calculation of $\Delta M$ and $\md$}\label{subsec:deltam}
Our hydrodynamical simulations enable us to calculate the binding energy of the material to the black hole $\dmde$. This function can be used to determine the feeding rate as a function of time through Kepler's third law,
\begin{equation}
\md = \frac{dM}{dE}\frac{dE}{dt} = \frac{2\pi}{3} \left(G\mh\right)^{2/3} \frac{dM}{dE} t^{-5/3}.
\end{equation}
For full disruptions, the entirety of the star's original mass is included in the calculation of $\dmde$, approximately half of which will have specific orbit energy $E > 0$ and is thus unbound from the black hole. For partial disruptions, the criteria for determining which material to include in the determination of $\dmde$ is less straightforward, as what will be accreted by the black hole is only the material that the star's gravity is unable to retain. As the star is on a parabolic orbit, the distance from the black hole changes rapidly as a function of time, and thus the star's Hill radius $a_{\rm H}(t) \equiv r (M_{\rm bound}(t)/\mh)^{1/3}$ is also time-dependent, introducing some ambiguity into the determination of the self-bound mass $M_{\rm bound}(t)$.

In principle, the distance of matter from the surviving star can be compared to $a_{\rm H} (t)$ to determine what mass is bound to the star. However, the continual reaccretion of matter means that the star is extended, non-spherical, and dynamically unrelaxed for many dynamical timescales, and thus the appropriate mass to use in the calculation of $a_{\rm H}$ is uncertain. To circumvent this, we choose an iterative energy-based approach that we find converges quickly to a solution. First, we calculate the material that remains bound to the star, where the initial reference point is taken to be at the location of the star's peak density, which has a velocity ${\bf v}_{\rm peak}$. The specific binding energy of material in a given cell is calculated as
\begin{equation}
E_{\ast,i} = \frac{1}{2}\left({\bf v}_{i} - {\bf v}_{\rm peak}\right)^{2} - \phi_{\ast}\label{eq:east},
\end{equation}
where $\phi_{\ast}$ is the gravitational self-potential as calculated by the multipole solver. The center of momentum ${\bf v}_{\rm cm}$ is then determined by summing over all mass elements for which $E_{\ast,i} < 0$
\begin{equation}
{\bf v}_{\rm cm} =\frac{ \sum_{E_{\ast,i} < 0} {\bf v}_{i} \rho_{i} V_{i}}{\sum_{E_{\ast,i} < 0} {\rho_{i} V_{i}}},
\end{equation}
where $\rho_{i}$ and $V_{i}$ are the cell density and volume. Equation (\ref{eq:east}) is then re-evaluated with ${\bf v}_{\rm peak}$ being replaced by ${\bf v}_{\rm cm}$. This process is repeated until ${\bf v}_{\rm cm}$ (and thus $M_{\rm bound}$) converges to a constant value. While this approach yields a value for $M_{\rm bound}$ in most cases, the question of whether an object is completely destroyed is somewhat complicated by the fact that the tidal force formally approaches zero as $\Delta r \rightarrow 0$, and thus there is always some material for which ${\bf v}_{i} = {\bf v}_{\rm cm}$, resulting in a infinitesimal, but non-zero $M_{\rm bound}$ even as the time since disruption $t - t_{\rm d} \rightarrow \infty$.

\begin{figure*}
\centering\includegraphics[width=0.9\linewidth,clip=true]{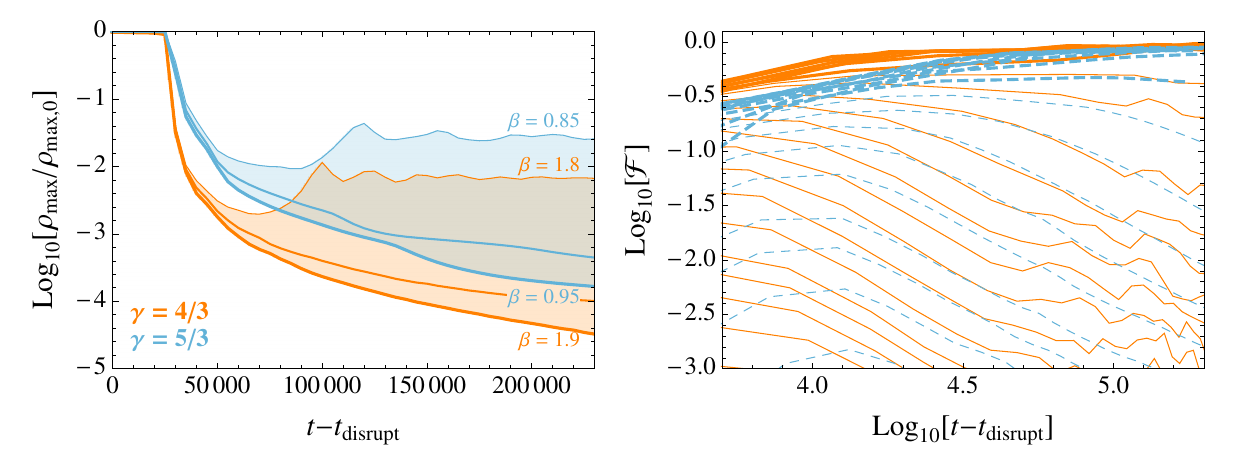}
\caption{Evolution of maximum density $\rho_{\max}$ and bound mass $M_{\rm bound}$ as a function of time since disruption. In the left panel, the evolution of the ratio of $\rho_{\max}$ to the original maximum density $\rho_{\max,0}$ is shown for six simulations (filled regions), three for $\gamma = 4/3$ (orange, solid lines) with $\beta$ = 0.85, 0.9, and 0.95, and three for $\gamma = 5/3$ (light blue, dashed lines) with $\beta$ = 1.8, 1.85, and 1.9. In the right panel, the parameter ${\cal F} \equiv |\dot{M}_{\rm bound} / M_{\rm bound}| (t-t_{\rm d})$ is shown for all simulations, demonstrating the convergence of the calculate $M_{\rm bound}$ for all the simulations presented in this work (see Section \ref{subsec:deltam} for details). When the value of this quantity is close to unity, $M_{\rm bound}$ is still changing by order unity over that timescale, indicating that the final bound mass cannot yet be determined at that $t$. The thick lines show simulations for which the star is considering to be destroyed after the encounter, whereas the thin lines show simulations for which a surviving core forms.}
\label{fig:convergence}
\end{figure*}

Figure \ref{fig:convergence} shows how the maximum density within a simulation $\rho_{\max}$ compares to the star's initial maximum density $\rho_{\max,0}$ for six simulations (three for $\gamma = 4/3$ and three for $\gamma = 5/3$). Two of the simulations shown for each $\gamma$ exhibit a continuous decrease in $\rho_{\max}$, showing no signs of recollapse, whereas the third simulation for each $\gamma$ shows an increase in density sometime after pericenter, eventually settling to a constant value as the collapsed object dynamically relaxes. As a check on the convergence of $M_{\rm bound}$ for all the simulations presented in this work, we compute the quantity ${\cal F} \equiv \left|\dot{M}_{\rm bound} / M_{\rm bound}\right| (t-t_{\rm d})$, which expresses the fractional change in $M_{\rm bound}$ since the time of disruption. Disruptions in which a self-bound core forms asymptote quickly to a constant $M_{\rm bound}$, and thus small values of ${\cal F}$, whereas disruptions in which $\rho_{\max}$ consistently decreases show ${\cal F} \sim 1$ at all times. The only disruptions in which the final core mass has not completely converged are the borderline survival cases (e.g. $\beta$ slightly less than $\beta_{\rm d}$). However, while the fractional error in $M_{\rm bound}$ is large for the borderline cases, the definition of $\Delta M = M_{\ast} - M_{\rm bound}$ means that the amount of mass lost from the star (and also the amount of mass bound to the black hole) is well-determined.

Once ${\bf v}_{\rm cm}$ has been determined, all material for which $E_{\ast,i} < 0$ is excluded, and the binding energy to the black hole $E$ is calculated. This data is then binned in $E$, the result of which is used to determine $\dmde$. The values of $\Delta M$ and $\md$ presented in the figures in the subsequent sections are all generated from snapshots that are produced at $t = 2.5 \times 10^{5}$ s after the start of each simulation (unless otherwise noted), or approximately 100 dynamical times after pericenter.

\section{Hydrodynamics of the tidal disruption of MS stars}\label{sec:results}

Many assumptions about the way partial and full disruptions work have never been tested beyond analytical approximations. Quantities that have been estimated include the time of return of the most bound material $t_{\rm most}$, the time of peak accretion rate $\tpeak$ and the magnitude of this rate $\mdpeak$, and the amount of mass bound both to the star and to the black hole after the encounter. Additionally, it has always been presumed that the late-time evolution of the fallback converges to the $t^{-5/3}$ decay law, whereas this is not necessarily true in partial disruptions where the surviving core may affect the binding energy of this material. We empirically measure these quantities from our calculations of $\dmde$, and find that while some of the commonly-held assumptions are reasonably accurate, many are not. Most of these assumptions arise from how the problem was originally formulated, in which the star's self-gravity is viewed as inconsequential, and only the spread in binding energy across the star at pericenter is relevant in determining the features of the resulting $\md$. We find that the star's self-gravity is critical in determining the resulting $\md$, even for encounters with pericenters that are many times deeper than the tidal radius.

\subsection{The Boundary Between Survival or Destruction}\label{subsec:survival}

A collection of non-interacting particles in the presence of a point mass potential will all follow Keplerian orbits, provided that no outside force acts upon them. This means that once both the star's gravity and pressure become unimportant at a time close to the star's closest approach to a black hole, the position and velocity of each mass element can be recorded, and the future orbits of each part of the debris stream can be determined. It has been presumed that this condition is satisfied at $\rt$, the distance at which the tidal force is greater than the self-gravitational force at the object's surface. This assumption is flawed in that the tidal radius as classically defined does not denote the distance at which the tidal force dominates self-gravity for any point within the star, but rather only at its surface. The conditions necessary for a polytrope to lose mass due to the presence of an external tidal force have been previously determined in the context of the Roche problem, which considers when the tidal force at the surface of an object exceeds the self-gravitational force in a circular orbit \citep{Aizenman:1968gn,Chandrasekhar:1969uj}. However, again, this limit only informs us as to when we expect the object to begin losing mass, and not the distance at which the object is completely destroyed. Additionally, the Roche limit is evaluated under the assumption of hydrostatic equilibrium, and presumes that the orbital velocity is equal to that of a circular orbit $v_{\rm c}$, resulting in a different dynamical response than for parabolic encounters in which the pericenter velocity is $\sqrt{2} v_{\rm c}$.

The question of whether a star survives depends not on the ability of tidal forces to remove some mass, but on whether these forces are overwhelming enough to disrupt the star's densest regions. Furthermore, even if a star experiences a seemingly complete disruption, the star may be capable of recollapse into a self-bound object after the encounter under the proper conditions. It has been shown that gamma-law equations of state stiffer than $\Gamma = 2$ can result in the recollapse of material within expanding thin streams for infinitesimally small masses \citep{Chandrasekhar:1961uk,Lee:2007em}. As stars are well-approximated by $\Gamma \leq 5/3$ equations of state, these instabilities are not expected to appear in stellar disruptions, and thus recollapse is not guaranteed for all $\beta$.

The affine model, as introduced in \citet{Carter:1985ti}, improved upon the initial estimates provided by the Roche approach by including the effects of the dynamical tide, but while this approach is able to evaluate the distance at which distortions become non-linear, it is not capable of determining the actual distance at which disruptions occur. Later, \cite{Diener:1995ui} extended the affine approach to calculate the critical impact parameter for full disruption $\beta_{\rm d}$, finding $\beta_{\rm d} = 1.12$ for $\gamma = 4/3$ and $\beta_{\rm d} = 0.67$ for $\gamma = 5/3$ polytropes, where $\beta_{\rm d}$ is the critical impact parameter at which  complete disruption ensues. More recently, the affine formalism was improved upon further by modeling the star as a nested set of ellipsoids, each of which respond dynamically to the external tidal field \citep{Ivanov:2001fva,Ivanov:2003dz}. While this model is the first analytical approach to provide estimates for $\Delta M$, the simplifying assumptions made regarding the treatment of self-gravity, pressure, and geometry does not guarantee that the true solution can be recovered via this approach.

\begin{figure}
\centering\includegraphics[width=\linewidth,clip=true]{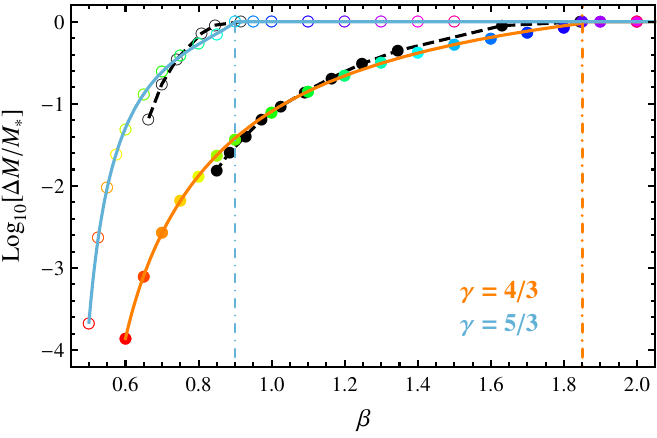}
\caption{Fits to $\Delta M$, with the fits to the $\gamma = 4/3$ models being shown by the solid colored circles, and fits to the $\gamma = 5/3$ models being shown by the open colored circles. Predictions of $\Delta M$ from \cite{Ivanov:2001fva} for both $\gamma = 4/3$ and $\gamma = 5/3$ are represented by the black symbols/curves. The color coding matches that of Figures \ref{fig:disrupt-beta-4-3} and \ref{fig:disrupt-beta-5-3}, with the impact parameters $\beta_{\rm d}$ beyond which stars are considered to be destroyed being denoted by the colored dot-dashed lines.}
\label{fig:mlost}
\end{figure}

\begin{figure*}
\centering\includegraphics[width=0.785\linewidth,clip=true]{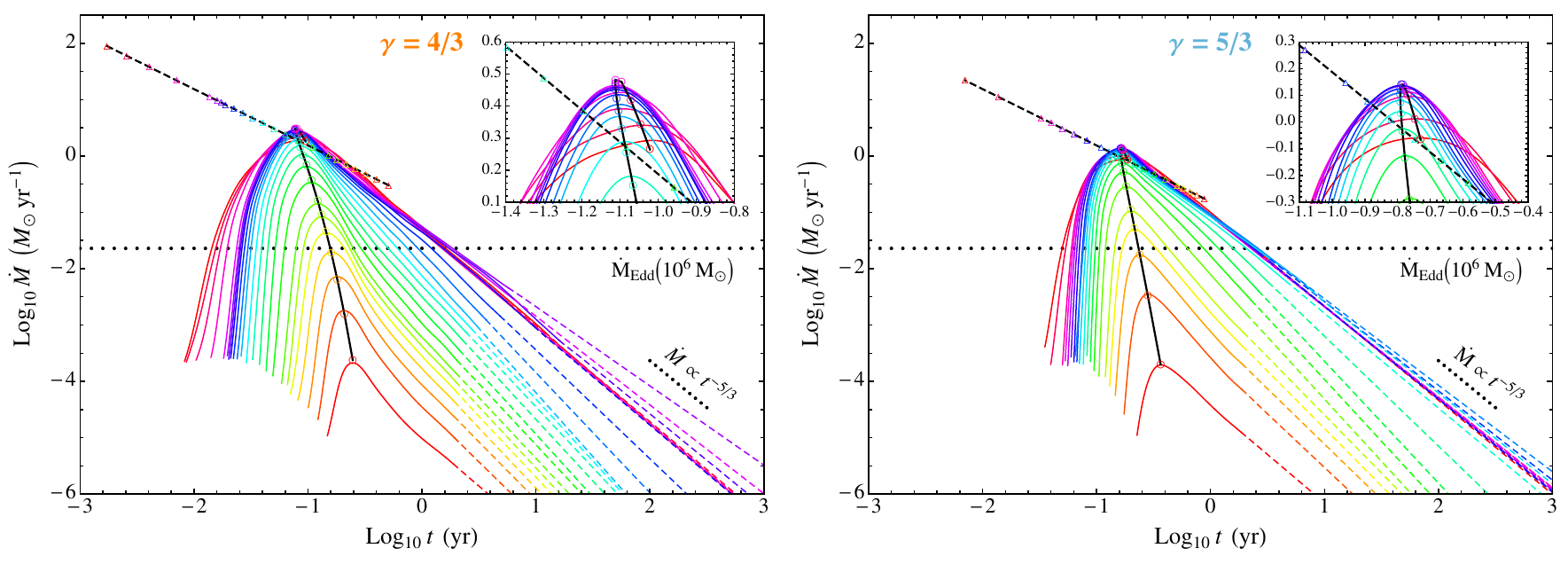}\includegraphics[width=0.215\linewidth,clip=true]{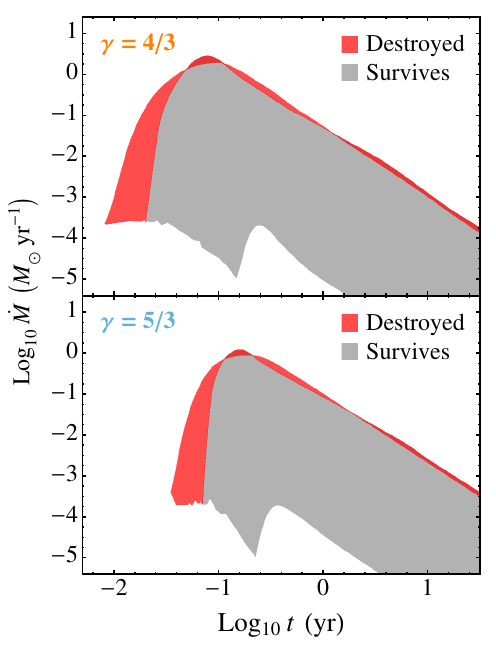}
\caption{Fallback accretion rate $\md$ onto a $10^{6} \msun$ black hole from the disruption of a 1 $\msun$ star as a function of $\gamma$ and $\beta$. The colored curves in the left two panels show $\md$, with the color of each curve corresponding to the color coding scheme presented in Figures \ref{fig:disrupt-beta-4-3} and \ref{fig:disrupt-beta-5-3}. The dashed portions of each curve are extrapolations based on the slope of the $\md$ immediately prior to the extrapolated region, which accounts for the fact that the late-time slope can only be determined exactly if the simulations are run for a prohibitive amount of time (see Figure \ref{fig:evol}). The dotted line shows the Eddington limit for a $10^{6} \msun$ black hole assuming an accretion efficiency $\epsilon = 0.1$. The open triangles connected by the gray dashed line show the peak fallback rate $\mdpeak$ and time of peak $\tpeak$ as predicted by the energy-freezing model, in which the period of the return of materials scales as $\beta^{3}$ \citep{Evans:1989jua,Ulmer:1999jja,Lodato:2009iba}. The open circles connected by the black line show the fits to $M_{\rm peak}$ and $\tpeak$ as given by equations \ref{eq:mdotpeak} and \ref{eq:tpeak} respectively. The right two panels shows the same data as the left two panels, with the filled regions showing the range of $\md$ curves resulting from full disruptions (red) and from disruptions in which the star survives (gray).}
\label{fig:accrate}
\end{figure*}

In Figure \ref{fig:mlost} we show the amount of mass lost $\Delta M = M_{\ast} - M_{\rm bound}$ (measured at the end of each simulation) as a function of $\beta$ for both $\gamma = 4/3$ and $\gamma = 5/3$, with comparisons to \citet{Ivanov:2001fva} shown in black. Remarkably, the model of \citeauthor{Ivanov:2001fva} comes quite close to predicting the critical $\beta$ value as measured by these simulations, despite the assumptions made, and is able to recover reasonable values for $\Delta M$, although the scaling between $\Delta M$ and $\beta$ is somewhat steeper than what is observed in the simulations. In particular, we find that stars can survive encounters for larger values of $\beta$ than the nested affine model predicts. We speculate that the method presented in \citeauthor{Ivanov:2001fva} could be extended to calculate $\md$ if the time at which each ellipsoid becomes unbound were recorded, which given the low computational burden of this approach could be used to perform more extensive parameter space studies.

We observe that while some stars appear initially to be completely destroyed, with their cores being disrupted along with their envelopes (Figures \ref{fig:disrupt-beta-4-3} and \ref{fig:disrupt-beta-5-3}), the debris stream can often recollapse many dynamical timescales after the encounter, resulting in a small yet self-bound remnant. The mass of the remnant that results is roughly equal to the amount of mass contained within a sphere centered at the recollapse point and with a radius equal to the cylindrical radius of the debris stream $S$.

For collapsing gaseous cylinders, spurious condensations as the result of the accumulation of numerical error may develop if the Jeans length is not properly resolved \citep{Truelove:1997bj}, with the source of that error being exacerbated by an inexact determination of the gravitational potential \citep{Jiang:2013ii}. \citeauthor{Truelove:1997bj} found that no spurious gravitational collapse occurs if the ratio $J$ of the grid scale to the Jeans length $\lambda_{\rm J} \equiv \sqrt{\pi c_{\rm s}^{2}/G\rho}$, where $c_{\rm s}$ is the sound speed and $\rho$ is the density, is always less than 0.25 in all grid cells at all times. In all of our simulations, the width of the debris stream is comparable to the star's initial size, and the resolution in the densest portion of the stream as it condenses is equal to the resolution used to resolve the original star ($\sim 50$ grid cells). Therefore, in the case in which a recollapse marginally occurs (i.e. $J/S \sim 1$), $J \simeq 0.02$, satisfying the Truelove criteria.

For $\gamma = 4/3$, we find that stars are destroyed for $\beta \geq \beta_{\rm d} = 1.85$, i.e. no self-bound stellar remnant is produced. To verify that we are adequately resolving the boundary between survival and destruction, we ran a single $\gamma = 4/3, \beta = 1.8$ simulation at double the linear resolution, and found a recollapse that results in a bound remnant of only a few percent of a solar mass, slightly smaller than what is found using our fiducial resolution. As the mass of the surviving star nears zero, the resolution requirements become progressively more restrictive, as even slight changes in the cylindrical density profile or gravitational potential can alter the time of recollapse, and thus the final bound mass. For $\gamma = 5/3$, we find that stars are destroyed for $\beta \geq \beta_{\rm d} = 0.9$.

Numerical challenges aside, the exact boundary between survival and destruction for real stars is likely to be slightly different than what is predicted here, as the central densities of stars on the MS depend on rotation, metallicity, and age \citep{Maeder:1974te,Wagner:1974fy}. Notably, our own Sun has a central density approximately twice that of the standard $\gamma = 4/3$ polytrope used to model it. This may allow the cores of somewhat evolved MS stars to survive for slightly larger values of $\beta$, although their gravitational influence is likely small as the helium-enriched cores of evolved MS stars are no larger than 10\% of the star's mass \citep{Schonberg:1942hh}.

\subsection{Characteristic features of $\md$}\label{sec:characteristic}

Figure \ref{fig:accrate} shows the family of $\md$ curves as a function of $\beta$ for both $\gamma = 4/3$ and $\gamma = 5/3$. Immediately evident is the strong dependence between $\mdpeak$ and $\beta$ for $\beta < \beta_{\rm d}$, and the similarity of the $\md$ curve family for $\beta \geq \beta_{\rm d}$. The result that deeper encounters do not produce more rapid flares is in direct conflict with the analytical prescription presented in \cite{Lodato:2009iba} (hereafter \lodato), in which the binding energy $dM/dE$ is equivalent to the spread in mass over distance (modulo a constant), $dM/dx$, at pericenter. In this model (hereafter referred to as the ``freezing model''), the binding energy is given by 
\begin{equation}
E = G \mh x / r_{\rm p}^{2},\label{eq:binen}
\end{equation}
and thus deeper encounters always result in faster-peaking transients. Because the binding energy $E \propto r_{\rm p}^{-2}$, the scaling between $\beta$ and $\tpeak$ is expected to be $\tpeak \propto \beta^{3}$ \citep{Ulmer:1999jja}.

We definitively find that this is not the case, as the two separate functional forms of the parametric pair $\left[\tpeak(\beta),\mdpeak(\beta)\right]$ indicate a separate set of assumptions are appropriate for the two cases $\beta < \beta_{\rm d}$ and $\beta > \beta_{\rm d}$, neither of which match the functional form advocated by \lodato (Figure \ref{fig:accrate}, triangles). For encounters in which $\beta < \beta_{\rm d}$, $\tpeak$ and $\mdpeak$ are approximately related to one another by a power law, with the best fit model having $\mdpeak \propto \tpeak^{-7.4}$ for $\gamma = 4/3$ and $\mdpeak \propto \tpeak^{-10.5}$ for $\gamma = 5/3$. The steepness of this relation means that the difference in $\tpeak$ is only a few tenths of a dex between an event in which $10^{-4} \msun$ is lost and a full disruption. For $\beta > \beta_{\rm d}$, the trend between $\tpeak$ and $\mdpeak$ reverses for increasing $\beta$, with deep encounters resulting in both slightly longer duration flares and slightly lower typical accretion rates.

For fully-disruptive encounters, we find that $\md$ varies little with increasing $\beta$. An assumption of the freezing model is that the distance at which the dynamics of the debris can be described by Kepler's laws is when the star is at pericenter. In fact, the star's self-gravity becomes unimportant {\it before} the star comes this close to the black hole for encounters where $\beta > \beta_{\rm d}$. This suggests that the binding energy distribution of the material should be determined shortly after the star crosses the full disruption radius $r_{\rm d} \equiv r_{\rm t} / \beta_{\rm d}$, and not at its closest approach, unless the encounter is grazing enough such that $r_{\rm p} < r_{\rm d}$.

\begin{figure}
\centering\includegraphics[width=\linewidth,clip=true]{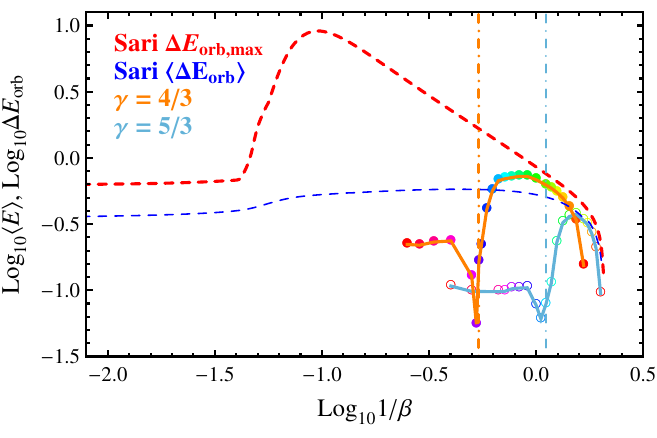}
\caption{Average spread of matter post-disruption as compared to the change in orbital energy for binary disruptions \citep{Sari:2010de}. The solid curves show $\langle E\rangle$ (the mass-averaged binding energy of the bound debris post-disruption) for both $\gamma = 4/3$ (light blue, open circles) and $\gamma = 5/3$ (orange, filled circles) stars, whereas the dashed curves show the change in orbital energy $\Delta E_{\rm orb}$ for prograde binary encounters, where we have presumed that each star has mass $M_{\odot}$. The impact parameters $\beta_{\rm d}$ beyond which stars are considered to be destroyed being denoted by the colored dot-dashed lines. The red dashed curve shows the maximum change in orbital energy $\Delta E_{\rm orb,\max}$ at a particular $\beta$, whereas the blue dashed curve shows $E_{\rm orb}$ averaged over binary phase. The binary disruption energies are scaled by $(G M_{\ast} / a) (M_{\rm h}/M_{\odot})^{1/3}$, where $a$ is the initial binary separation, whereas the stellar disruption curves are scaled by $(G M_{\ast} / R_{\ast}) (M_{\rm h}/M_{\odot})^{1/3}$.}
\label{fig:spread}
\end{figure}

\begin{figure*}
\centering\includegraphics[width=0.9\linewidth,clip=true]{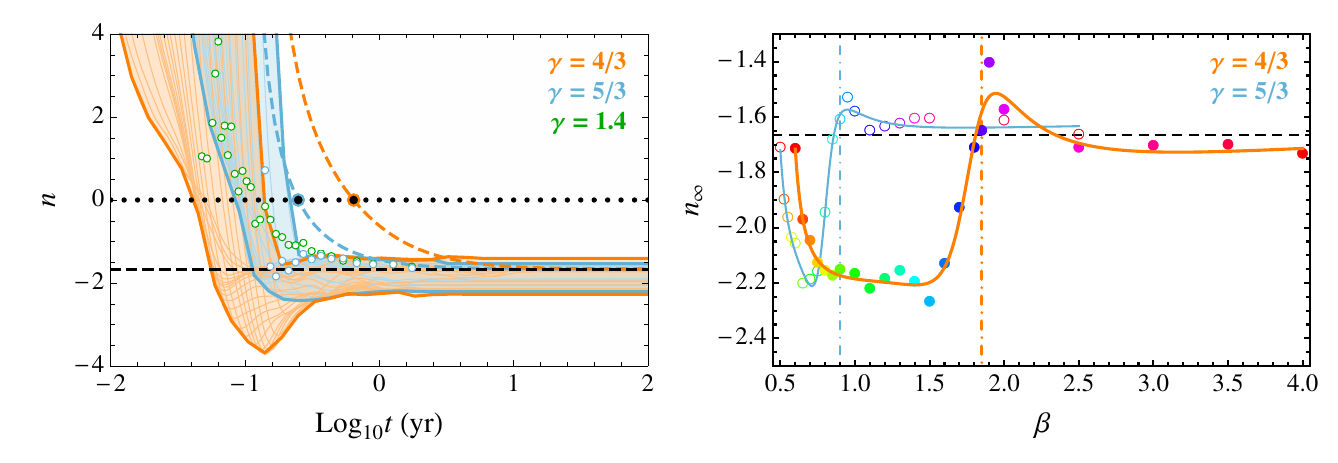}
\caption{The left panel shows the power-law index $n$, with orange corresponding to $\gamma = 4/3$ stars and light blue corresponding to $\gamma = 5/3$ stars. The filled regions show the convex hull of $n$ over all values of $\beta$; the lightly-weighted curves within the filled regions represent individual simulations for specific values of $\beta$. The horizontal dotted and dashed black curves show $n = 0$ (i.e. the peak of the accretion rate) and $n = -5/3$ (the canonical value for constant $\dmde$), respectively. The dashed colored curves are produced using the analytical formulae of \cite{Lodato:2009iba} for $\beta = 1$ encounters for both values of $\gamma$. For 1 $\msun$ stars, the analytical formulae predict a faster rise to peak for $\gamma = 4/3$ (orange-bordered point) than for $\gamma = 5/3$ (blue-bordered point), and fail to reproduce the steeper power law index that is found shortly after peak for $\gamma = 4/3$. The open colored circles show the numerical results of \cite{Lodato:2009iba} for $\beta = 1$, where we include their $\gamma = 1.4$ case (dark green) in addition to $\gamma = 5/3$ and note that their simulations set $\Gamma = \gamma$. The right panel shows the asymptotic power-law index $\ninf$ as a function of $\beta$, with the color coding scheme identical to that of Figures \ref{fig:disrupt-beta-4-3} and \ref{fig:disrupt-beta-5-3}, where the filled circles show $\ninf$ for $\gamma = 4/3$, and the open circles for $\gamma = 5/3$. A best fit for both values of $\gamma$ are shown by the solid colored curves, and the impact parameters $\beta_{\rm d}$ beyond which stars are considered to be destroyed are denoted by the colored dot-dashed lines.}
\label{fig:n}
\end{figure*}

This can be understood by considering the local reaction time of each layer of the star's structure as compared to the passage timescale. The dynamical timescale for a particular layer is $\tau_{\rm dyn} \simeq \sqrt{1/G \bar{\rho}_{\rm x}}$, which is approximately equal to the time between when the star is at a distance where the tidal force is capable of removing that layer and the time of pericenter,
\begin{eqnarray}
\tau_{\rm tidal} &=& r_{{\rm t},x}/v_{{\rm t},x} \simeq \sqrt{\frac{r_{{\rm t},x}^{3}}{G \mh}} = \sqrt{\frac{x^{3}}{G M_{x}}}\\
&\simeq& \sqrt{1/G\bar{\rho}_{x}},
\end{eqnarray}
where the subscript $x$ refers to quantities defined by the mass interior to $x$. Thus, regardless of the distance at which the tidal force begins to dominate the self-gravitational force, material is removed from the star at or near the full disruption radius $r_{\rm d}$. This means that the effective radius that should be used in the denominator of equation (\ref{eq:binen}) is $r_{\rm eff} =\max(r_{\rm d}, r_{\rm p})$. However, as the degree of balance between the tidal and self-gravitational forces continuously evolves over the encounter, the actual radius at which mass is removed can be larger or smaller than $r_{\rm eff}$, and thus the relationship between $E$ and $x$ is more complicated than outlined here.

Additionally, while the binding energy is effectively frozen-in once the star crosses $r_{\rm d}$, the assumption that the orbital energy can be reliably recorded at this point is only valid if the pressure gradient that develops within the star during maximum compression is not large enough to affect $\dmde$. As shown in \cite{Carter:1983tz}, the pressure component of the Lagrangian does build significantly shortly after pericenter, and eventually dominates the tidal component for sufficiently deep encounters. However, while this build-up can lead to the production of shocks whose breakouts may be observable as short X-ray transients \citep{Guillochon:2009di}, we find that the gradient of pressure within the orbital plane primarily acts to redistribute the most highly-bound material for ($t \lesssim \tpeak$), and not the material that determines the behavior of the decay phase 
(see Figure 3 of \citeauthor{Guillochon:2009di} and Figure 5 of \lodato). The tangible effect of this pressure build-up on the shape of $\md$ is the spreading of some material that would have otherwise accreted at $\tpeak$ to more highly-bound orbits, thus reducing the rate of accretion at peak, shifting $\tpeak$ to later times, and leading to an increased feeding rate at early times.

This behavior is analogous with what is found for binary star disruptions, in which the change in orbital energy $\Delta E_{\rm orb}$ of the stars is independent of the impact parameter for sufficiently deep encounters \citep{Sari:2010de}. In Figure \ref{fig:spread}, we compare $\Delta E_{\rm orb}$ calculated by \citeauthor{Sari:2010de} for binary disruptions to the mass-averaged spread in the binding energy $\langle E\rangle$ of the material that becomes bound to the black hole. As is found in binary disruption calculations, the change in energy initially increases with increasing $\beta$, then a transition point is reached where the binary's gravity (or star's gravity, in our case) no longer affects the dynamics, and finally the change in energy approaches a constant. A single star disruption and the disruption of a binary system are conceptually quite similar. A full disruption is analogous to an equal-mass binary disruption with separation distance $a \sim R_{\ast}$, where the mass of each ``star'' is equal to the mass liberated from each Lagrange point, $M_{1} = M_{2} = M_{\ast} / 2$. A partial disruption is analogous to an unequal mass ``trinary'' system, in which the three masses correspond to the surviving self-bound core with mass $M_{\ast} - \Delta M$, and the bound/unbound debris streams with mass $\Delta M/2$, all with initial separation $a \sim R_{\ast}$. As the results presented in \citeauthor{Sari:2010de} are independent of mass, the normalized $\Delta E$ for disruptions still map closely to those seen for binary disruptions, despite the variance in mass of the two (or three) interacting objects.

A caveat in our comparison to binary systems is that binaries can have an arbitrary orientation upon arrival at pericenter. This leads to an increase in the potential maximum energy change, and the $\beta$ value at which it occurs. This comes as the result of stars in a binary being able to come arbitrarily close to one another during an encounter for favorable binary phases at pericenter ($\Delta E_{\rm orb,\max}$ in Figure \ref{fig:spread}), which permits them to interact gravitationally for longer. In effect, the binary can become ``denser,'' decreasing the size of its effective tidal radius. For a stellar disruption in which the star is not initially rotating, the mass interior to a given radius cannot increase in the same way, and thus its self-gravity ceases to be important interior to its original tidal radius. This results in the near--constant spread in energy as described above, and is visually evident from simulation snapshots (Figures \ref{fig:disrupt-beta-4-3} and \ref{fig:disrupt-beta-5-3}), in which the debris distributions are almost identical for $\beta \gtrsim \beta_{\rm d}$. We speculate that for rapidly rotating stars that this same effect that can yield large $\Delta E$ for certain binary phases may also apply, as stellar rotation can permit stars to penetrate more deeply before $\dmde$ is set \citep{Stone:2012ul}.

Figure \ref{fig:n} shows both the power-law slope $n(t)$ over the full $\md$ curve (left panel), and the asymptotic power-law slopes $\ninf$ (right panel), as produced by our disruption simulations. The behavior of the curves is more complicated than what is implied by the freezing model, in which $n \geq -5/3$ for all $t$. The qualitative behavior of the $\md$ curves can be characterized by three phases: A {\it rise} phase, in which $n > 0$, a {\it drop} phase, in which $n < 0$ (and potentially even < -5/3), and an {\it asymptotic} phase, in which $\ninf \equiv n(t \rightarrow \infty)$ assumes a constant value. The rise phase is somewhat similar to what is predicted by the freezing model, although the evolution of $n$ is somewhat more rapid within the simulations. The drop phase exhibits particularly steep downward slopes for $\gamma = 4/3$ stars, with $n \sim -4$ shortly after peak, but $n$ for $\gamma = 5/3$ stars is closer to the predicted asymptotic value. Despite the disagreement between our simulations and the analytical model presented in \lodato, we do find we reproduce the simulation results of LKP for $\gamma = 5/3$ stars at the same $\beta$ (Figures \ref{fig:n} and \ref{fig:structure}).

The discrepancy between the simulation results and the prediction of the freezing model can be understood by considering what material becomes bound to the black hole in the case that the star is not completely destroyed. The binding energy $E$ of material a distance $x$ from the center of the star is $\propto x / \max(r_{\rm d}, r_{\rm p})^{2} \propto x \beta^{2}$ (assuming $x \ll r_{\rm p}$). The value of $x$ that corresponds to the material that determines the asymptotic behavior of $\md$ can be estimated by considering the deepest point within the star during the encounter in which tidal forces are capable of overcoming the star's self-gravitational force,  the exact functional form of which  is dependent upon the hydrodynamical response of the star during the encounter. While this functional form can only truly be determined through hydrodynamical simulation, it is clear that the effective $x$ must decrease with increasing $\beta$, as the tidal forces remove an ever-increasing fraction of the star's mass. This implies that the scaling between $E$ and $\beta$ must be weaker than $\beta^{2}$, and thus $\tpeak$ should show less evolution for progressively deeper, but not completely disruptive encounters.

The asymptotic phase exhibits a more complicated behavior that depends on $\beta$, and shows four distinct behaviors depending on the depth of the encounter for both $\gamma = 4/3$ and $\gamma = 5/3$ stars. For extremely grazing encounters in which a small fraction of the star's mass is lost, $\ninf \simeq -5/3$. In these encounters, all of the mass is removed near pericenter, resulting in an energy spread that only depends on $x$, the distance to the star's center of mass (in agreement with the freezing model).

\begin{figure}
\centering\includegraphics[width=0.8\linewidth,clip=true]{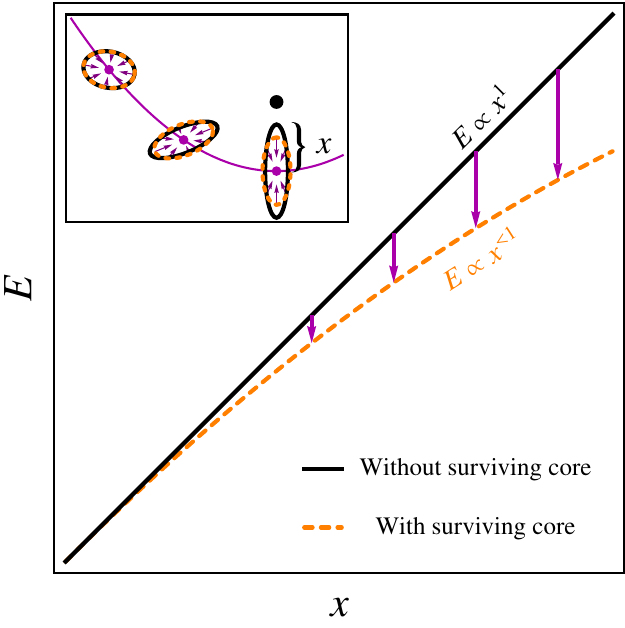}
\caption{Cartoon showing the gravitational effect of a surviving core on the dynamics of material that is removed from the star during a partially-disruptive encounter. The inset diagram in the upper left demonstrates how the restoring force provided by a surviving core can alter the structure of the outer layers. For encounters in which the core plays little role, the binding energy to the black hole $E$ scales linearly with $x$ (black solid curve), the distance from the star's center of mass \citep{Lodato:2009iba}. If a core survives the encounter, its gravity prevents material from moving as quickly away from the star, resulting in a weaker relationship between $E$ and $x$ (orange dashed curve). This consequently results in a more-steeply declining $\md$. If the core itself is close to destruction, its gravitational influence is minimal, and the canonical $\md \propto t^{-5/3}$ decay law is recovered.}
\label{fig:dedx}
\end{figure}

\begin{figure*}[t]
\centering
$\vcenter{\hbox{\includegraphics[width=0.43\linewidth,clip=true]{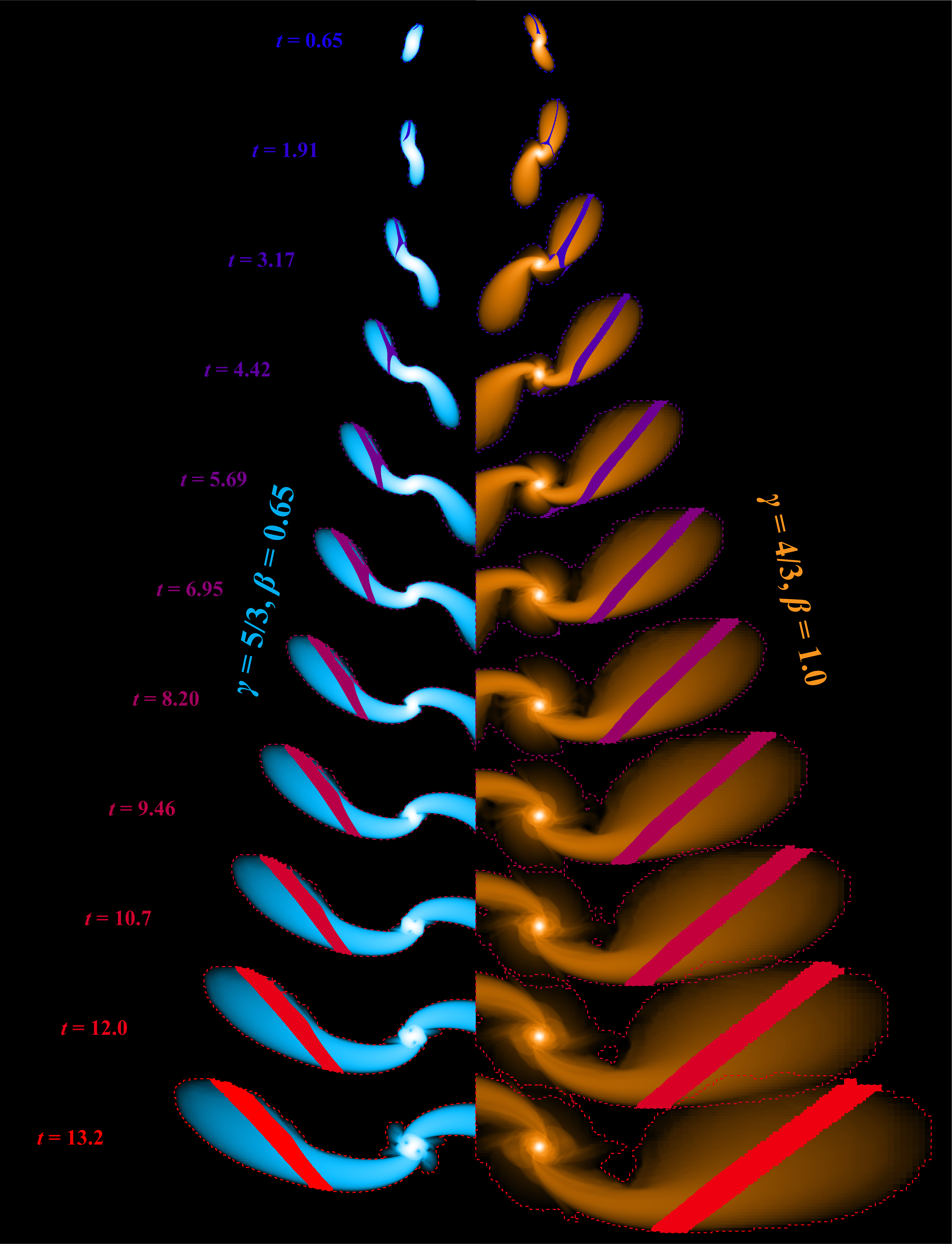}}}$
$\vcenter{\hbox{\includegraphics[width=0.52\linewidth,clip=true]{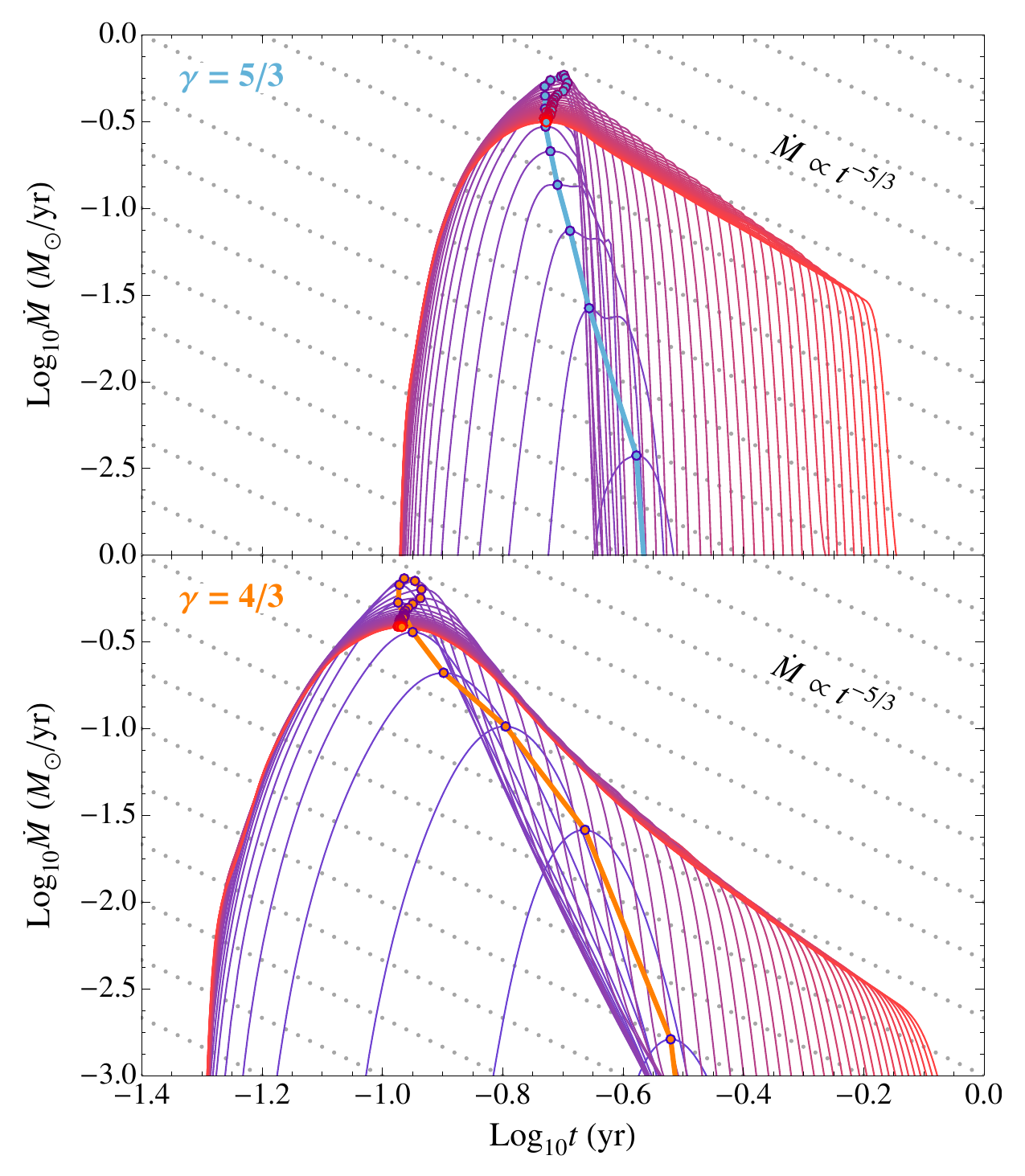}}}$
\caption{The left panel shows a collage from two simulations of the regions that contain the mass that contributes to $\mdpeak(t)$, super-imposed over the density distribution of the each snapshot. The fill and outline color denotes the time of the snapshot after pericenter, with blue corresponding to $t = t_{\rm p}$, and red corresponding to \(t = t_{\rm p} + 2 \times 10^{4}\). Snapshots are shown from two different disruption simulations that have similar values of $\Delta M$, with light blue showing the disruption of a $\gamma = 5/3$ star for $\beta = 0.65$, and orange showing the disruption of a $\gamma = 4/3$ star for $\beta = 1.0$. Each fill region shows the material that contributes to the part of $\md$ that is within 90\% of $\mdpeak(t)$. The right panel shows the values of $\md$ derived from these two simulations (one curve per snapshot), with the colors of each curve corresponding to the color of the fill regions of each snapshot. The solid lines correspond to $\gamma = 5/3$, and the dashed lines corresponding to $\gamma = 4/3$, with the thick light blue and orange lines being fitted to $\mdpeak(t)$.}
\label{fig:gamma-comp}
\end{figure*}

When a significant fraction of the star's mass is removed in an encounter, $\ninf$ steepens to values as large as $\simeq -2.2$. This behavior arises from the influence of the star's core (Figure \ref{fig:dedx}). As the outermost layers of the star are removed prior to pericenter, the core is able to partially counter the black hole's tidal force, keeping material closer to the star's core, and thus reducing the effective $x$ at which the material is no longer strongly affected by the core's gravity. This results in a sub-linear relationship between $E$ and $x$. As $E \propto t^{-2/3}$, and $E \propto x^{m}$, where $m \leq 1$, the resulting asymptotic power-law is
\begin{equation}
\ninf = \frac{2}{3m}-\frac{7}{3}
\end{equation}
where $\ninf = -5/3$ is recovered for the standard linear relationship. If $m < 0$, this implies that $E$ actually decreases with $x$, and thus the most bound material would initially lie interior to the least bound. As the most bound material would have to cross beyond the least bound, we expect that any $m < 0$ relationship would be quickly flattened to at least $m = 0$ by pressure gradients, resulting in a limit on the asymptotic slope of $\ninf \geq -7/3$.

For disruptions that are just deep enough to destroy the star (i.e. $\beta = \beta_{\rm d}$), $\ninf$ can be somewhat less steep than -5/3. This implies that $m > 1$; the relationship between $E$ and $x$ is super-linear. For this borderline case only, the release of material that eventually composes the decay tail of $\md$ is moderated by the slow shrinkage of the stellar core, which is not fully destroyed until after the star has passed pericenter. For these encounters, $r$ and $x$ are somewhat dependent, with the material being released at the smallest $x$ being launched at large $r$, and thus the quantity $x/r^{2}$ can be $\propto x^{> 1}$.

Finally, for deep encounters, $\ninf$ again seems to be consistent with -5/3. Unlike the borderline case where a core persists long after pericenter, here the core is rapidly destroyed, and the energy is again set at a fixed $r$. As there is no core to resist the tidal force, the energy spread is simply given by the spread in potential energy across the star,  \`a la the freezing model.

\subsection{The Influence of Stellar Structure}\label{sec:structure}

A seemingly counter-intuitive result in the context of the freezing model is the fact that less-centrally concentrated stars, which have more mass at larger radii, result in transient events that peak at later times than their more centrally concentrated brethren, even for events that yield the same $\Delta M$. As shown in Section \ref{subsec:survival}, the distance at which total disruption occurs is significantly deeper for centrally concentrated stars, which may explain some of the discrepancy. Consider what happens to a star in the approach to pericenter for two extreme cases:  A case in which most of the star's mass is concentrated at its center, and a case in which the star has near-constant density. In the centrally-concentrated example, the star's outer layers will find that their dynamics are partly determined by the tidal force at early times, but also partly determined by the core, which remains initially undisturbed (Figure \ref{fig:dedx}). The influence of any surviving core on the dynamics of the matter can thus affect the final binding energy $E$.

\begin{figure*}[t]
\centering\includegraphics[width=\linewidth,clip=true]{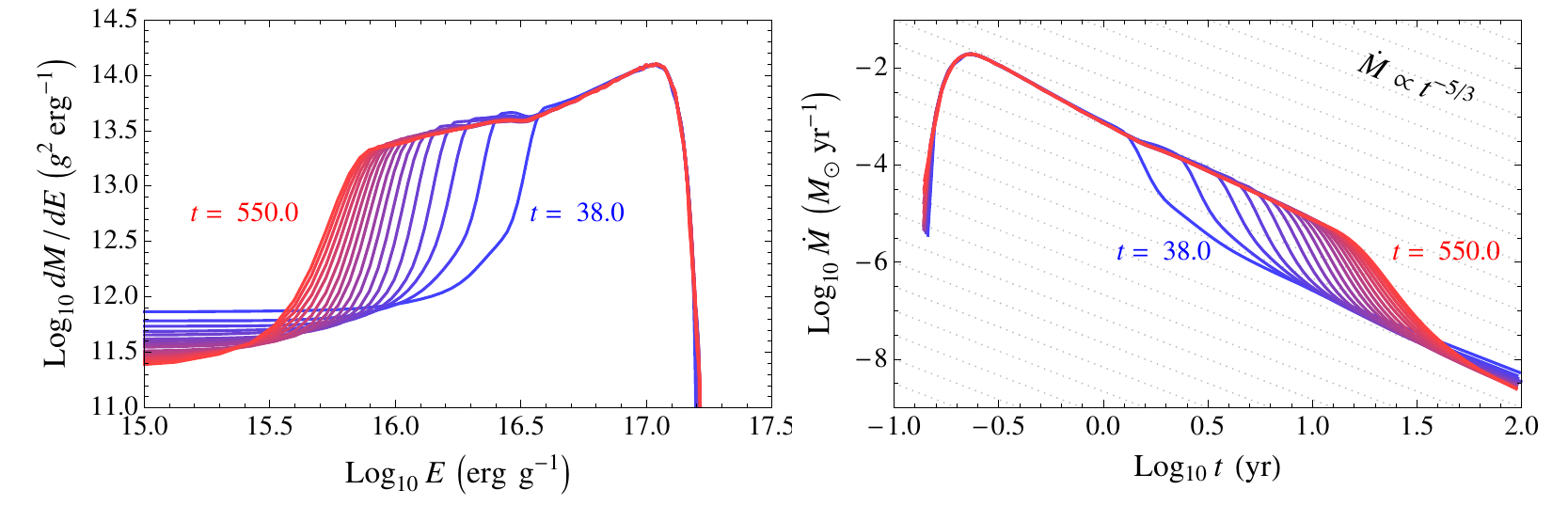}
\caption{Evolution of $\dmde$ (left panel) and $\md$ (right panel) for the disruption of a star with polytropic index $\gamma = 5/3$ and impact parameter $\beta = 0.55$. The measured mass distribution in both plots is shown as a function of time relative to the time of pericenter, with the blue curves indicating early times and the red curves showing late times. The right-hand plot is overlaid on a series of dotted gray lines showing the fiducial $\md \propto t^{-5/3}$ evolution. While the early-time $\md$ can be determined shortly after the disruption, the rate of fallback still evolves for $t \gtrsim 10$ yr even after the simulation has been allowed to run for many hundreds of dynamical timescales.}
\label{fig:evol}
\end{figure*}

The left panel of Figure \ref{fig:gamma-comp} shows that one of the fundamental assumptions of the freezing model, that the binding energy $E$ and the distance from the star's core $x$ are linearly related, is not correct, and matter that contributes to a particular $E$ is drawn from a range in $x$ that spans nearly the entire star. In Figure \ref{fig:gamma-comp} we compare two disruptions with nearly identical $\Delta M$ for $\gamma = 4/3$ and $\gamma = 5/3$. The left panel shows the time evolution of the material that determines the peak of $\md$, with the contour colors corresponding to the same times after pericenter. The filled contours show the regions that contribute to $\mdpeak$ within each snapshot. The expectation under the freezing approximation would be that all mass that possesses a given energy $E$ comes from a cylindrical cross-section of the star, but as illustrated in  Figure \ref{fig:gamma-comp}  the geometry of the debris that contributes to $\mdpeak$  is clearly not cylindrical. The right panel shows the $\md$ that correspond to these snapshots. For the earliest snapshots (light blue-filled contours), the material in the $\gamma = 5/3$ simulation appears to have a head-start over the centrally-concentrated case, despite the fact that the $\gamma = 4/3$ encounter is 50\% deeper ($\beta = 1.0$ for $\gamma = 5/3$ versus $\beta = 0.65$ for $\gamma = 4/3$). However, as the encounter progresses, the peak of $\md$ for $\gamma = 4/3$ moves to progressively earlier times relative to $\gamma = 5/3$, eventually settling to a value that results in a faster transient.

This implies that the binding energy of these layers relative to the black hole should not be recorded assuming the star has preserved its original spherical shape and size. While less material is positioned near the black hole when comparing the centrally-concentrated case to the constant density case, the core of the star continues to interact with the debris, and effectively ``carries'' material to larger $E$ before $E$ has been fixed. For constant density stars, the tidal force and the self-gravitational force scale to the same power in $x$, and the core is disrupted at approximately the same time as the outer layers, which is consequently why these stars are destroyed at a distance that more closely matches the classical Roche result. In this case, the assumption that $E$ can be determined by considering the star's original size and shape is more appropriate, as little stellar material is carried closer to the black hole, as is found in the centrally-concentrated case.

\subsection{Long-term evolution of $\md$}

\begin{figure*}[htb]
\centering\includegraphics[width=0.95\linewidth,clip=true]{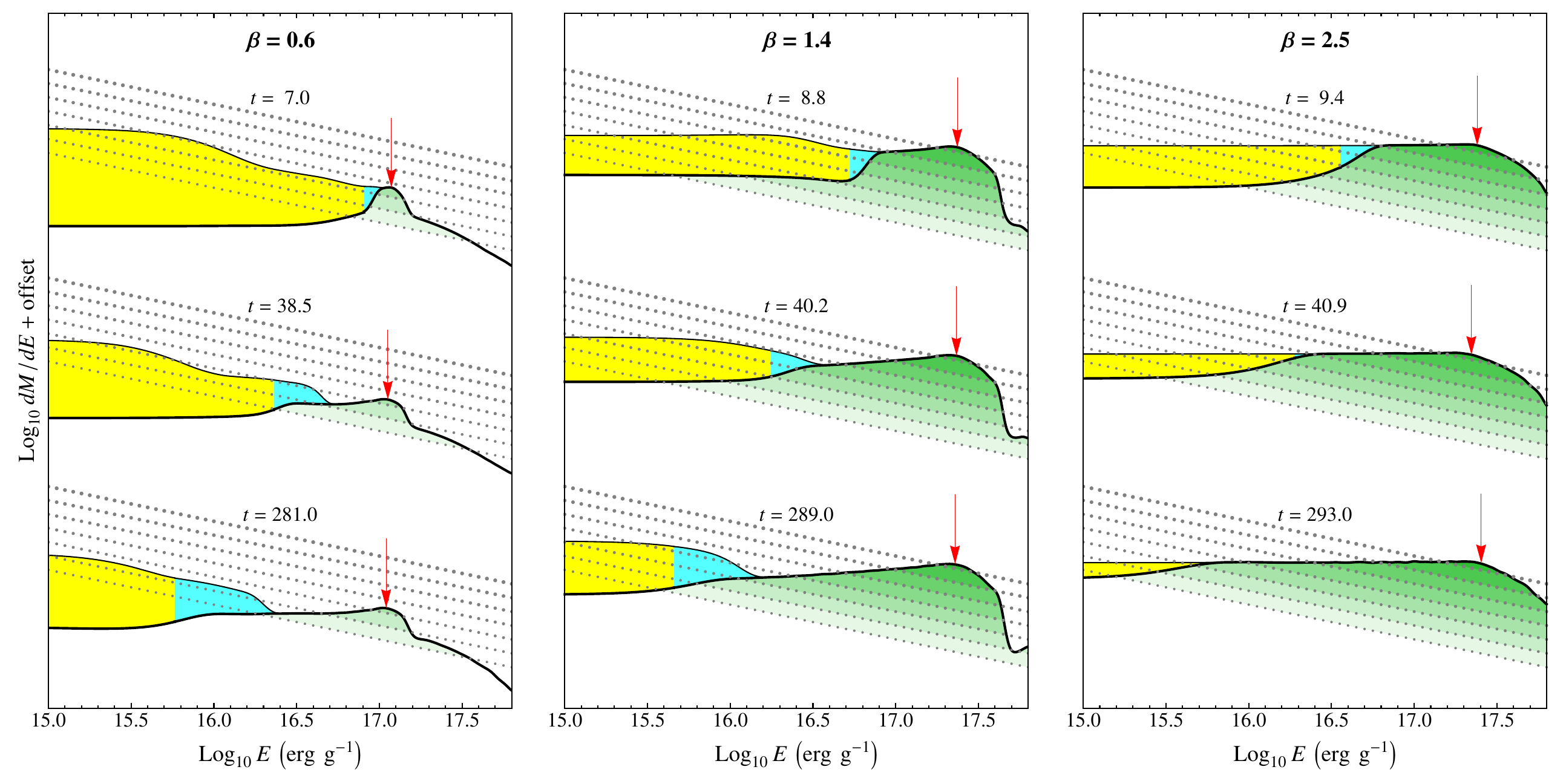}
\caption{Distribution of mass as a function of binding energy $E$ for three different simulations of a $\gamma = 4/3$ disruption for $\beta = 0.6$, 1.4, and 2.5. Shown in each panel is a sequence of mass distributions in time, with the evolution in time progressing from top to bottom, where the yellow regions show all material that remains bound to the black hole, and the green regions show material bound to the black hole but {\it not} bound to the star. The cyan region shows material that is bound to the star, but whose semi-major axis larger than the distance defined by the surviving core's time-dependent Hill sphere, $a_{\rm H}(t)$. The green regions are separated by the gray dotted curves into sub-regions where the resulting accretion rate $\md$ exceeds, from lowest to highest, $10^{-6}$, $10^{-5}$, $10^{-4}$, $10^{-3}$, $10^{-2}$, $10^{-1}$, and 1 $\msun$/yr. The red arrows show the location within the material bound to the black hole which determines the peak accretion rate $\mdpeak$.}
\label{fig:dmde}
\end{figure*}

While the peak of the accretion rate is determined within tens of stellar dynamical timescales, the tail of $\md$ continues to evolve for hundreds of dynamical timescales. Most notably, a large ``cavity'' in $\md$ is present for accretion times $t > \tpeak$ (Figure \ref{fig:evol}), which gradually fills from left to right as additional material in the tidal debris tails satisfies the simple energy criteria applied to determine if matter is bound to the black hole. An example of the long-term evolution of $\md$ arising from this interaction is shown in Figure \ref{fig:evol}, where $\md$ is not determined for $t \gtrsim 10$ years until 550 dynamical timescales after the disruption. As the star recedes from the black hole, the evolution of $\md$ slows (as indicated by the decreasing space between the curves in Figure \ref{fig:evol}), implying that progressively longer simulation times are required to determine $\md$ much beyond $\tpeak$. 

The cavity arises from the exclusion of material within the debris stream that remains bound to the stellar core after the encounter, with the evolution coming as a result of the continued interaction between either the stellar core (if the star survived the encounter) or a mildly self-gravitating debris stream (if it did not) and the black hole. An examination of the pressure of the debris tails reveals that the debris is free-streaming, even in the vicinity of the Hill sphere. In other words, the pressure gradients present within the stream are small enough to be incapable of modifying the material's trajectory. Thus, the interaction between the stream, black hole, and surviving core is purely gravitational.

In Figure \ref{fig:dmde} we show $\dmde$ for disruptions of a $\gamma = 4/3$ star for three values of $\beta$. As material that is considered to be bound to the star (yellow) crosses the time-dependent Hill sphere, it becomes bound to the black hole (green). We find that there is always some mass in the vicinity of the time-dependent Hill sphere $a_{\rm H}(t)$ (cyan) for encounters in which a core survives, whereas full disruptions do not show this behavior, mostly because the self-bound mass shrinks drastically as progressively less material satisfies the criteria for being self-bound.

As the star moves away from the black hole, the Hill radius grows, but by a rate that is a factor $(M_\ast/\mh)^{1/3}$ smaller than the rate at which the star recedes from the black hole. This implies that any material that retains a positive velocity relative to the surviving core after the encounter has the potential to be removed from the star, even if it is technically bound to the core (i.e. $v^{2} < 2 G M_{\rm core} / r$) at an earlier epoch. Much of the observed evolution of $\dmde$ may be due to our definition for what is considered to be ``bound'' to the surviving core after the encounter (Section \ref{sec:method}). Our calculation presumes that the energy budget of material with respect to the star is sufficient to determine what inevitably remains bound to the star; in reality the question of whether a given particle remains bound or not amounts to solving the restricted elliptical three-body problem, for which no closed-form solution exists. The Jacobi constant, which has a fixed value in the restricted circular three-body problem and can be used to determine the zones within which a particle of a given initial position and velocity can occupy \citep{Murray:1999th}, is not constant once the orbit is non-circular \citep{Hamilton:1992eh}.

\begin{figure*}
\centering\includegraphics[width=0.9\linewidth,clip=true]{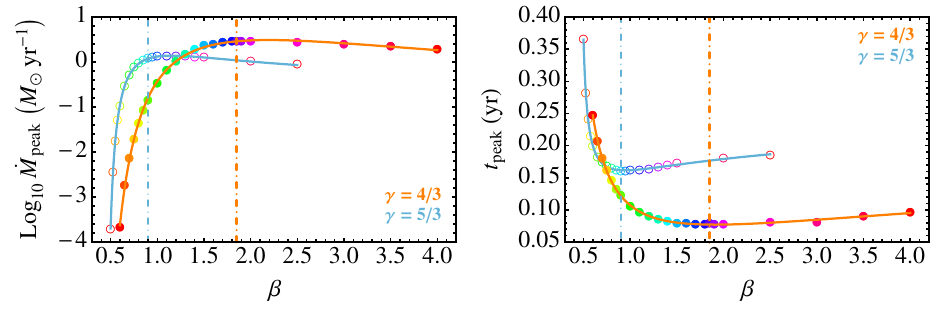}
\caption{Fits to $\mdpeak$ and $\tpeak$, with the fits to the $\gamma = 4/3$ models being shown by the solid colored circles, and fits to the $\gamma = 5/3$ models being shown by the open colored circles. The color coding matches that of Figures \ref{fig:disrupt-beta-4-3} and \ref{fig:disrupt-beta-5-3}, with the impact parameters $\beta_{\rm d}$ beyond which stars are considered to be destroyed being denoted by the colored dot-dashed lines.}
\label{fig:peak}
\end{figure*}

However, while the energy balance approach may not be capable of immediately determining the mass that will eventually become bound to the black hole, the distribution only remains uncertain for material that is accreted far beyond $t_{\rm peak}$. In the limit that $t \rightarrow \infty$, the distance of the star to the black hole increases as $t^{2/3}$, and thus $E$ for material leaving the Hill sphere is $G M_{\rm h} a_{\rm h} / r^{2} \propto t^{-2/3}$. This explains the observed slowing of the evolution of $\md$.

The material that fills in the cavity assumes a distribution in $E$ that is not entirely flat, resulting in a fallback rate that scales to a power slightly steeper than the canonical $t^{-5/3}$. This energy distribution is likely set near pericenter, where the pressure component is comparable to the tidal component. As the star recedes from the black hole, the pressure component of the force decreases more quickly than the tidal component \citep{Kochanek:1994bn}, and thus the debris is expected to evolve purely gravitationally. However, the conditions under which material is launched across the time-dependent Hill sphere may depend somewhat on the pressure gradient, no matter how small, as the net gravitational force is zero \citep{Lubow:1975bf}.

\begin{figure*}[t]
\centering\includegraphics[width=0.9\linewidth,clip=true]{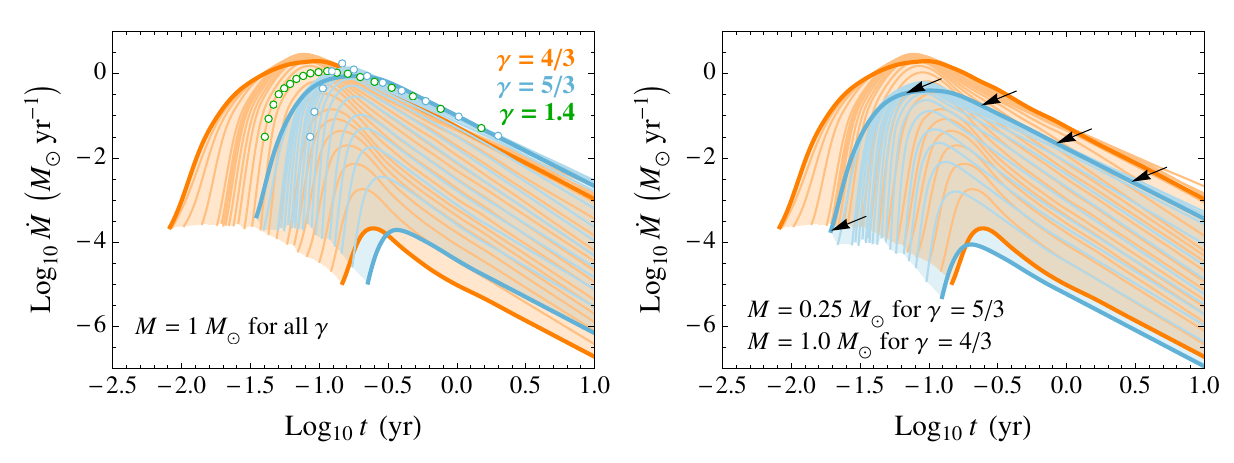}
\caption{Comparison of the families of $\md$ curves for $\gamma = 4/3$ (orange) and $\gamma = 5/3$ (light blue). The left panel shows $\md$ derived from the simulations presented here as solid lines, assuming that the stars for both $\gamma$ have mass $M = 1 \msun$. The open circles show the numerical results of \citep{Lodato:2009iba} for $\beta = 1$ and $\gamma = 1.4$ (dark green) and $\gamma = 5/3$ (light blue). The right panel shows the shift in $\md$ (black arrows) if the mass and radius of star that is expected to have a structure described by $\gamma = 5/3$ is taken into account \citep{Tout:1996waa}.}
\label{fig:structure}
\end{figure*}

\section{Discussion}\label{sec:discussion}
The results of our extensive parameter study produce a number of unexpected trends as compared to the predictions presented by previous work. In the previous section we attempted to explain the observed scalings, and how these features arise as a result of the interaction between the black hole and a potentially surviving stellar core. In what follows, we explain how these newly discovered features can be used to constrain the type of disrupted star and how it was disrupted.

\subsection{Can $\gamma$ and $\beta$ be determined a posteriori?}\label{sec:aposteriori}

Given our predicted $\md$, is it possible to determine either the stellar structure or the impact parameter from the light curve produced by a tidal disruption event? The conversion efficiency between mass accreted by the black hole and the light emitted is somewhat uncertain, and depends on factors such as the black hole's spin and how the accretion rate compares to $\dot{M}_{\rm Edd}$ \citep{Ulmer:1999jja,Beloborodov:1999wb,Strubbe:2009ek,Lodato:2010ic}. However, as the efficiency cannot be larger than unity, and as flares are typically observed in the decay phase, we can only place lower limits on the amount of mass accreted by a black hole to produce a given flare \citep{Gezari:2008iv}. Thus, at the very least, our predicted $\Delta M$ (Figure \ref{fig:mlost}) can be used to exclude events for $\beta$ less than some critical value, given the mass of the star.

Figures \ref{fig:n} and \ref{fig:peak} present four additional quantities that enable us to classify tidal disruptions based on the properties of observed tidal disruption flares. Two of these quantities, $\mdpeak$ and $\tpeak$, are only available to us for flares in which the peak of the accretion rate is clearly observed \citep{Gezari:2012fk}, but both $n(t)$ and $\ninf$ are measurable for flares that are observed long after peak \citep{Komossa:1999uy,Komossa:2004dr,Gezari:2006gd,Gezari:2008iv,Cappelluti:2009jl,vanVelzen:2011gz,Cenko:2012fg}. If the mass of the black hole is known with some certainty, one may be able to infer both $M_{\ast}$ and $\beta$ by simply measuring $\mdpeak$ and $\tpeak$ and comparing to our resultant $\md$, which at first glance appear to form distinct sequences for $\gamma = 4/3$ and $\gamma = 5/3$ stars (Figure \ref{fig:structure}, left panel). However, this is only true assuming that centrally concentrated stars have the same mass and radius as stars of near constant density. The transition from stars that are well-modeled by a $\gamma = 4/3$ polytrope to a $\gamma = 5/3$ polytrope is also accompanied by a decrease in radius such that all stars with mass $0.25 M_\odot < M_\ast < M_\odot$ have the same central density \citep{Kippenhahn:1990tm}. Adjusting the radii and mass of our $\gamma = 5/3$ models to the mass and radius of a $0.25 M_\odot$ star \citep{Tout:1996waa}, we find that the sequence of $\md$ functions for $1.0 M_\odot$ and $0.25 M_\odot$ stars lie on top of each other (Figure \ref{fig:structure}, right panel), making the determination of the disrupted mass of a star somewhat degenerate with its structure.

This motivates us to look for other features of $\md$ that may uniquely identify either $\gamma$ or $\beta$. If we consider the power-law of the rate of decline $n$ after peak, we find that there is a distinguishing feature between $\gamma = 5/3$ and $\gamma = 4/3$ models at $\sim 0.5$ dex after $\tpeak$. Whereas $\gamma = 5/3$ stars quickly converge to $n \simeq -5/3$, $\gamma = 4/3$ models show a characteristic drop, with $n$ being as large as -4 for some encounters (Figure \ref{fig:n}, left panel). This feature is most prominent for intermediate $\beta$ in which $\sim 50\%$ of the star's mass is removed during the encounter, and represents the strong influence of the dense stellar core, which acts to drag material deeper within the black hole's potential before tidal forces are capable of removing it.

In addition to being more centrally-concentrated to begin with, an additional component that likely contributes to this observed drop is the adiabatic response of the surviving core. For $\gamma = 5/3$ stars, the removal of mass results in the inflation of the star, whereas $\gamma = 4/3$ exhibit the opposite behavior, shrinking dramatically in response to the loss of mass \citep{Hjellming:1987ci}. This enhances the core's influence during the encounter in the phase where the core's mass is changing, slowing the reduction in the core's effective gravity, and thus pulling even more matter to higher binding energies. The recently observed flare PS1-10jh presented in \cite{Gezari:2012fk} shows a clear drop in the accretion rate with respect to the canonical $t^{-5/3}$ decline rate expected from the freezing model. In the freezing model, it is impossible to produce a decline feature steeper than $t^{-5/3}$ within any part of $\md$, as we explained in Section \ref{sec:characteristic}. As many tidal disruption flares may show this characteristic drop in $\md$, a clearly-resolved peak can be used to compare to the subsequent decay phase for a precise determination of $n(t)$.

For events in which the peak is not clearly observed, and for which the signal-to-noise is too small to permit an accurate determination of $n(t)$, the asymptotic slope $\ninf$ of $\md$ can still provide additional information about the star that was disrupted. As shown in the right panel of Figure \ref{fig:n}, $\ninf$ can be used to distinguish between partial and full disruptions. The fact that $\ninf$ assumes values that are significantly steeper than -5/3 may indicate that additional tidal disruption flares have been found observationally, but subsequently discarded and/or ignored due to the mismatch between the measured $n$ and $-5/3$ \citep{vanVelzen:2011gz}. This implies that some supernovae that have been observed at the centers of galaxies may in fact be misidentified partial tidal disruptions.

\subsection{Future work}

As found in previous work \citep{Faber:2005be,Guillochon:2011be}, there is a change in surviving star's orbital energy after the encounter, with the change in energy being comparable to the star's initial self-binding energy. This change in energy, combined with the star's initial orbital energy, leads to a shift in the entire $\dmde$ distribution, which can affect the fallback of material for $E \sim \Delta E_{\rm orb}$, or for $t \gtrsim \mh R_{\ast}^{3/2} G^{-1/2} M_{\ast}^{-3/2} \sim 100$ years, given that $\Delta E_{\rm orb} \sim G M_{\ast}/R_{\ast}$. As the star's initial orbital energy may not be zero and can be comparable to $\Delta E$ itself, and thus the final binding energy of the star depends on its initial orbital energy, we have presented our $\dmde$ and $\md$ curves with this change in energy removed. As a result, our plots show the fallback rate that would be expected if the final star were to remain on a parabolic trajectory, as our initial conditions assume. While these kicks that are typically of the order of star's own escape velocity may be important in determining the further fate of the star and whether it will suffer additional disruptions, they are not expected to affect the first century of a flare's evolution, of which only the first few years are accessible to currently available transient surveys. 

Even if $\md$ is directly related to the properties of the star being disrupted, the luminosity of the accretion disk $L$ may not directly follow $\md$. The primary factors that affect the link between $\md$ and the bolometric $L$ are the viscous evolution of the disk and the size of the disk \citep{RamirezRuiz:2009gw}, although other processes may strongly affect the amount of light observed in a single band, especially in the optical/UV where dust extinction can play a vital role. Disk viscosity can only affect $L$ for $t \lesssim \tau_{\rm visc}$, in which its primary affect is to delay emission at early times. However, once $t > \tau_{\rm visc}$, $L$ is expected to track $\md$ closely. As the material is delivered to the disk at $r \simeq r_{\rm p}$, the ratio of $\tau_{\rm visc}$ to $\tpeak$ is
\begin{align}
\frac{\tau_{\rm visc}}{\tpeak} = 3.2 &\times 10^{-2} \beta^{-3} \left(\frac{B_\gamma\left(\beta\right)}{0.1}\right)^{-1} \times\nonumber\\
&\left(\frac{\mh}{10^{6} \msun}\right)^{-1/2}
\left(\frac{\alpha}{0.1}\right)^{-1} \left(\frac{M_\ast}{M_\odot}\right)^{1/2},
\end{align}
where $B_\gamma$ is a fitted parameter derived from our simulations, ($B_\gamma \sim 0.1$ for most $\beta$, see Appendix \ref{sec:appendix}) and $\alpha$ is the parameterized $\alpha$-disk scaling coefficient, where we have taken the scale-height ratio $h/r = 1$. If $\tau_{\rm visc}/t_{\rm peak} \gtrsim 1$, the accretion is spread over longer timescales, resulting in a power-law decay index $n = -1.1$ \citep{Cannizzo:1990hw}. This may affect the light curve shape in the earliest phases of the fallback (prior to peak) where $t \ll \tau_{\rm visc}$, and thus the early evolution of $L(t)$ may not follow the functional forms of $\md$ presented here for $t \ll t_{\rm peak}$. But as observations of tidal disruption flares in the decay phase seem to be consistent with the canonical $n = -5/3$ decay law, it is clear that $\md$ and $L(t)$ must be closely coupled on year-long timescales.

An ingredient that the set of simulations presented in this paper do not include is the inclusion of general relativistic effects, which become important for very deeply penetrating encounters. Qualitatively, for both spinning and non-spinning black holes, general relativity is expected to result in more mass loss and a spreading of mass in $\dmde$ as compared to Newtonian encounters, as its primary effect is to bend the star's path such that it spends a larger fraction of time near the black hole where tidal forces are important \citep{Luminet:1985wz,Kobayashi:2004kq}. The only numerical provenance for how the metric may affect the feeding rate comes from low-resolution simulations performed by \cite{Laguna:1993cf}, which find a slight increase in $\mdpeak$ for increasing $\beta$, but much less than the predicted $\beta^{3}$ scaling. If the black hole has non-zero spin, the resulting $\dmde$ depends on the orientation of the star's angular momentum vector as compared to the black hole's spin vector \citep{Haas:2012ci}.

A spinning black hole permits deeper encounters that don't result in the star being immediately swallowed \citep{Kesden:2012cn}, provided that the two angular momentum vectors are aligned, and also should affect the final binding energy distribution, with co- and counter-rotational encounters resulting in smaller and larger $\Delta M$, respectively \citep{Diener:1997kw,Ivanov:2006fe,Kesden:2012kv}. However, as the fraction of disruptions in which non-Newtonian metrics can affect the dynamics is $\sim r_{\rm s}/r_{\rm t}$, which is $\sim 5\%$ for a $10^{6} \msun$ black hole and $\sim 20\%$ for a $10^{7} \msun$ black hole, the majority of tidal disruption events are well-represented by a Newtonian approximation to the black hole's gravity.

Lastly, the absence of hydrogen in spectra taken of the tidal disruption event PS1-10jh \citep{Gezari:2012fk} strongly suggests that the disruption of stars that are not on the MS may contribute significantly to the overall rate of tidal disruption. As we show, the structure of the star that is disrupted is clearly imprinted upon $\md$, providing valuable additional information that can be used to distinguish between candidate disruption victims. We explore the disruption of post-MS stars in a companion paper using a method similar to what is presented here \citep{MacLeod:2012cd}.

The discovery of flaring black hole candidates in nearby galaxies  will continue to elucidate the demography of the AGN population \citep{DeColle:2012bq}. Whereas AGN are supplied by a steady stream of fuel for hundreds or even thousands of years, tidal disruptions offer a unique opportunity to study a single black hole under a set of conditions that change over a range of timescales. There are, of course, rapidly varying stellar-mass black hole candidates in X-ray binaries within our own Galaxy. But for SMBHs, tidal disruption events offer the firmest hope of studying the evolution of their accretion disks for a wide range of mass accretion rates and feeding timescales.  The simulations and resultant $\md$ curves presented here are crucial for determining the properties of the black hole itself, as an incomplete model of a stellar disruption can result in much uncertainty in how the black hole converts matter into light. For a disruption with a well-resolved light curve, our models permit a significant reduction of the number of potential combinations of star and black hole properties, enabling a better characterization of SMBHs and the dense stellar clusters that surround them.

\acknowledgments We have benefited from many useful discussions with S.~Gezari, M.~Macleod, M.~C.~Miller, S.~Liu, R.~O'Leary, F.~Rasio, M.~Rees and C.~Thompson. We thank the anonymous referee for constructive corrections and suggestions. The software used in this work was in part developed by the DOE-supported ASCI/Alliance Center for Astrophysical Thermonuclear Flashes at the University of Chicago. Computations were performed on the UCSC Pleiades and Laozi computer clusters, and the NASA Pleiades computer cluster. We acknowledge support from the David and Lucille Packard Foundation, NSF grants PHY-0503584 and ST-0847563  and the NASA Earth and Space Science Fellowship (JG).

\appendix
\section{Fitting parameters}\label{sec:appendix}
For convenience, we have calculated fitting parameters for four characteristic quantities: The peak accretion rate $\mdpeak$, the time of peak accretion $\tpeak$, the amount of mass lost by the star $\Delta M$, and the asymptotic decay power-law index $\ninf$. These parameters can be used to constrain observed tidal disruption events based on measurable characteristics of their light curves (see Section \ref{sec:aposteriori}):
\begin{align}
\mdpeak &= A_{\gamma}\left(\frac{\mh}{10^{6} \msun}\right)^{-1/2} \left(\frac{\ms}{\msun}\right)^{2} \left(\frac{\rs}{R_{\odot}}\right)^{-3/2} \;\msun/{\rm yr}\label{eq:mdotpeak}\\
\tpeak &= B_{\gamma}\left(\frac{\mh}{10^{6} \msun}\right)^{1/2} \left(\frac{\ms}{\msun}\right)^{-1} \left(\frac{\rs}{R_{\odot}}\right)^{3/2} \;{\rm yr}\label{eq:tpeak}\\
\Delta M &= C_{\gamma} \ms\\
\ninf &= D_{\gamma}.
\end{align}
In these expressions are four functions of $\beta$ alone: $A_{\gamma}$, $B_{\gamma}$, $C_{\gamma}$, and $D_{\gamma}$. The forms of these functions are derived by fitting rational functions to the outputs produced by the numerical simulations presented in this paper. These functions are derived separately for two polytropic $\gamma$, $\gamma = 4/3$ and $\gamma = 5/3$, which are appropriate for high- and low-mass main sequence stars, respectively.

\begin{align}
A_{5/3} &= \exp\left[\frac{10.253-17.380 \beta +5.9988 \beta ^2}{1-0.46573 \beta -4.5066 \beta ^2}\right],&0.5 \leq \beta \leq 2.5\\
A_{4/3} &= \exp\left[\frac{27.261-27.516 \beta +3.8716 \beta ^2}{1-3.2605 \beta -1.3865 \beta ^2}\right],&0.6 \leq \beta \leq 4.0\\
B_{5/3} &= \frac{-0.30908+1.1804 \sqrt{\beta }-1.1764 \beta }{1+1.3089 \sqrt{\beta }-4.1940 \beta },&0.5 \leq \beta \leq 2.5\\
B_{4/3} &= \frac{-0.38670+0.57291 \sqrt{\beta }-0.31231 \beta }{1-1.2744 \sqrt{\beta }-0.90053 \beta },&0.6 \leq \beta \leq 4.0\\
C_{5/3} &= \exp\left[\frac{3.1647-6.3777\beta+3.1797\beta^2}{1-3.4137\beta+2.4616\beta^2}\right],&0.5 \leq \beta \leq 0.9\\
C_{4/3} &= \exp\left[\frac{12.996-31.149 \beta +12.865 \beta ^2}{1-5.3232 \beta +6.4262 \beta ^2}\right],&0.6 \leq \beta \leq 1.85\\
D_{5/3} &= \frac{-0.93653+11.109 \beta -38.161 \beta ^2+50.418 \beta ^3-22.965 \beta ^4}{1-8.6394 \beta +26.012 \beta ^2-32.383 \beta ^3+14.350 \beta ^4},&0.5 \leq \beta \leq 2.5\\
D_{4/3} &= \frac{-2.7332+6.9465 \beta -3.2743 \beta ^2-0.84659 \beta ^3+0.56254 \beta ^4}{1-2.3585 \beta +0.47593 \beta ^2+0.96280 \beta ^3-0.37996 \beta ^4}.&0.6 \leq \beta \leq 4.0
\end{align}

\bibliographystyle{apj}
\bibliography{apj-jour,/Users/james/Dropbox/library}

\end{document}